\documentclass[11pt, a4paper]{article}
\pdfoutput=1

\usepackage{jheppub}

\usepackage[T1]{fontenc} 

\usepackage{pgf}
\usepackage{longtable}
\usepackage{color,url}
\usepackage{colortbl}
\usepackage{multirow}
\usepackage{verbatim}
\usepackage[utf8]{inputenc}
\usepackage[colorlinks=true
,urlcolor=blue
,anchorcolor=blue
,citecolor=blue
,filecolor=blue
,linkcolor=blue
,menucolor=blue
,linktocpage=true
,pdfproducer=medialab
,pdfa=true
]{hyperref}

\usepackage{epsfig,psfrag,rotating,soul}
\usepackage{rotfloat}
\usepackage{epsf}
\usepackage{graphics}
\usepackage{color}
\newcommand{\RDp}{\ensuremath{ R_{D^{(*)}}} }
\newcommand{\RKp}{\ensuremath{ R_{K^{(*)}}} }
\newcommand{\Rk}{\ensuremath{R_{K}}}

\newcommand{\GeV}{\ensuremath{\,{\rm GeV}}}
\newcommand{\Clq}{\ensuremath{ C_{\ell q} } }
\newcommand{\Clqt}{\ensuremath{ C_{\ell q(3)} } }
\newcommand{\Clqo}{\ensuremath{ C_{\ell q(1)} } }
\newcommand{\CneNP}{\ensuremath{C_9^{\rm{NP}\, e}}}
\newcommand{\CteNP}{\ensuremath{C_{10}^{\rm{NP}\, e}}}
\newcommand{\CnmuNP}{\ensuremath{C_9^{\rm{NP}\, \mu}}}
\newcommand{\CtmuNP}{\ensuremath{C_{10}^{\rm{NP}\, \mu}}}

\title{Using Machine Learning techniques in phenomenological
    studies on flavour physics}

\author[a,b,c]{J. Alda,}
\author[d]{J.~Guasch,}
\author[a,b]{S.~Pe{\~n}aranda}

\affiliation[a]{Departamento de F{\'\i}sica Te{\'o}rica, Facultad de Ciencias,\\
Universidad de Zaragoza, Pedro Cerbuna 12,  E-50009 Zaragoza, Spain}
\affiliation[b]{Centro de Astropart{\'\i}culas y F{\'\i}sica de Altas Energ{\'\i}as (CAPA), 
Universidad de Zaragoza, Zaragoza, Spain}
\affiliation[c]{Dipartamento di Fisica e Astronomia ``G. Galilei'', Universit{\`a} degli
Studi di Padova e Istituto Nazionale Fisica Nucleare, Sezione di Padova,
I-35131 Padova, Italy}
\affiliation[d]{Deptartament de F{\'\i}sica Qu{\`a}ntica i
  Astrof{\'\i}sica and Institut de Ci{\`e}ncies del Cosmos (ICCUB),\\
Universitat de Barcelona, Mart{\'\i} i Franqu{\`e}s 1, E-08028 Barcelona, Catalonia, Spain}

\emailAdd{jalda@unizar.es}
\emailAdd{jaume.guasch@ub.edu}
\emailAdd{siannah@unizar.es}

\abstract{An updated analysis of New Physics violating Lepton Flavour
Universality, by using the Standard Model Effective Field Lagrangian with
semileptonic dimension six operators
at $\Lambda = 1\,\mathrm{TeV}$ is presented. We perform a global fit, by discussing the
relevance of the mixing in the first generation.
We use for the first time in this context a Montecarlo analysis to
extract the confidence intervals and correlations between observables. Our
results show that machine learning, made jointly with the SHAP values,
constitute a suitable strategy to use in this kind of analysis.}

\begin{document} 
\maketitle
\flushbottom

\section{Introduction}
\label{sec:intro}

The experimental measurement of Lepton Flavour Universality
Violating (LFUV) processes in $B$ meson decays, in tension with the
Standard Model (SM) predictions, would represent a clear sign
for physics beyond the SM. For the $b  \to s \ell^+ \ell^- $ processes,
observables such as the ratios of branching fractions $\RKp$,
\begin{equation}
\RKp = \frac{\mathrm{BR}(B\to K^{(*)} \mu^+ \mu^- )}{\mathrm{BR}(B\to K^{(*)} e^+ e^- )}\ ,
\end{equation}
provides evidence of LFUV and are of particular interest because
much of the theoretical uncertainty cancels in the ratio. It is well known
that in the SM, as a consequence of Lepton Flavour Universality (LFU),
$R_K = R_{K^*} = 1$ with uncertainties of the order of
$1\%$~\cite{Hiller:2003js,Bordone:2016gaq}.
However, the latest
experimental results from LHCb, in the specified regions of
$q^2$ di-lepton invariant mass, are:
\begin{align}
R_K^{[1.1, 6]} = 0.846^{+0.042}_{-0.039}{}^{+0.013}_{-0.012}\, \qquad &\mbox{\cite{Aaij:2021vac} }\nonumber \\
R_{K^*}^{[0.045,1.1]} = 0.66^{+0.11}_{-0.07}\pm 0.03 \qquad\qquad
R_{K^*}^{[1.1,6]} = 0.69^{+0.11}_{-0.07}\pm 0.05\ . \qquad
&\mbox{\cite{Aaij:2017vbb} }
\end{align}
Clearly, the results for the compatibility of the individual measurements with respect to the SM
predictions depend of the $q^2$ di-lepton invariant mass region, being 
of $3.1\, \sigma$ for the $R_K$ ratio, $2.3\, \sigma$ for the
$R_{K^*}$ ratio in the low-$q^2$ region and $2.4\, \sigma$ in the
central-$q^2$ region. The Belle collaboration has also reported
experimental results for the \RKp ratios~\cite{Abdesselam:2019lab,Abdesselam:2019wac}, 
although with less precision than the LHCb measurements.

Other targets of flavour violating processes are the $b\to c \ell \nu$
transitions. The ratios of branching fractions
$\RDp^\ell$ and $\RDp^\mu$, defined by, 
\begin{equation}
\RDp^\ell = \frac{\mathrm{BR}(B \to D^{(*)} \tau \bar{\nu}_\tau ) }{[\mathrm{BR}(B \to D^{(*)} e \bar{\nu}_e) + \mathrm{BR}(B \to D^{(*)} \mu \bar{\nu}_\mu)]/2}\ ,
\end{equation}
and
\begin{equation}
\RDp^\mu = \frac{\mathrm{BR}(B \to D^{(*)} \tau \bar{\nu}_\tau ) }{ \mathrm{BR}(B \to D^{(*)} \mu \bar{\nu}_\mu)}\ ,
\end{equation}
also exhibit sizeable deviations from their predicted SM values~\cite{Amhis:2019ckw},
\begin{equation}
R_D^{\ell\ \mathrm{SM}} = 0.299 \pm 0.003,\qquad\qquad
R_{D^*}^{\ell\ \mathrm{SM}} = R_{D^*}^{\mu\ \mathrm{SM}} = 0.258 \pm 0.005.
\end{equation}
Their measurements at BaBar~\cite{Lees:2012xj}, Belle~\cite{Abdesselam:2019dgh} and 
LHCb~\cite{Aaij:2017uff} experiments are larger than the SM
prediction. By assuming universality in the lighter 
leptons, the world average of the 
experimental values for the \RDp ratios, as obtained by the Heavy
Flavour Averaging Group (HFLAV), are~\cite{Amhis:2019ckw}
\begin{equation}
R_D^\mathrm{ave} = 0.340 \pm 0.027 \pm 0.013,\qquad\qquad
R_{D^*}^\mathrm{ave} = 0.295 \pm 0.011 \pm 0.008.
\end{equation}
These values imply a $1.4\,\sigma$ discrepancy with the SM predictions
for $R_D$, and $2.5\,\sigma$ for $R_{D^*}$. When combined
together, including their correlation, the excess is $3.08\,\sigma$.

There exist other observables displaying some discrepancies with
SM predictions even when larger theoretical uncertainties are taken into
account~\cite{ATLAS:2018cur,CMS:2014xfa,LHCb:2015wdu,ATLAS:2018gqc}.
It is clear than when investigating the implications of the experimental measurements in flavour
physics observables, a global fit should be considered.
Several global fits can be found in the literature (see, for
example~\cite{Capdevila:2017bsm,Celis:2017doq,Alok:2017sui,Camargo-Molina:2018cwu,Alda:2018mfy,Datta:2019zca,Aebischer:2019mlg,Aoude:2020dwv,Marzocca:2021azj,Alda:2020okk,Alda:2021ruz}
and references therein). We
have recently done a global fit to the updated experimental
information in~\cite{Alda:2020okk,Alda:2021ruz}, where an extensive list of
references to previous analyses is included.

From the theoretical point of view, Effective Field Theory is one of the most
widely used tools to study any possible New Physics (NP)
contribution. The effective Hamiltonian approach
allows us to perform a model-independent analysis of NP effects.
In this paper, we consider the Standard Model Effective Field Theory
(SMEFT) Lagrangian and we perform a global fit to the
Wilson coefficients using the packages
\texttt{flavio} v2.3~\cite{Straub:2018kue} and
\texttt{smelli} v2.3~\cite{Aebischer:2018iyb} (as described in details
in section~\ref{sec:Fits}).
The global fit includes the $\RKp$ and $\RDp$ observables,
the electroweak precision
observables; $W$ and $Z$ decay widths and branching ratios to leptons,
superallowed nuclear $\beta$ decays, all the available
experimental data for the related $b\to s\ell^+\ell^-$ observables;
i.e. the $b\to s \mu^+\mu^-$ observables (including the optimized angular
observable $P_5'$ and the branching ratio of $B_s \to \mu^+\mu^-$,
as well as all the
available data on angular observables in $B\to K^{(*)} \mu^+ \mu^-$ decay),
the relevant data related to $B\to K^{(*)} e^+ e^-$ decays, and also
the angular observables measured in different bins for
$B_{s} \to \phi \mu\mu$ decays. Finally,
the $b \to s \nu \bar{\nu}$ observables are also included in the global
fit. 

Because the Gaussian approximation to characterize the fit is not
successful, we will use for the first time in this context  a Montecarlo
analysis to extract the confidence intervals and other relevant
statistics, and we explicitly show that machine learning, taking jointly
with the SHAP (SHAPley Additive exPlanation) values, constitute a
suitable strategy to use in this analysis. 

The rest of this work is organized as follows: Section~\ref{sec:EFT} presents
a brief summary of the Effective Field Theory used to describe possible
NP contributions to $B$ decays observables. 
We then discuss in section~\ref{sec:Fits} the details of the global
fits performed, introducing the
phenomenological scenarios that we used in the analysis and presenting our
results. We found that the Gaussian approximation is not suitable to
characterize the fit and, therefore, in order to
extract the confidence intervals and other relevant statistics, we use
a Montecarlo analysis that is described in section~\ref{sec:MC}.
The agreement of the results obtained by the Machine Learning Montecarlo algorithm
that we have proposed and the ones obtained by using the Renormalization Group 
equations is also included in this section. 
Section~\ref{sec:LQ} includes a discussion of the phenomenological implications
of our analysis in leptoquark models. 
The conclusions are presented in section~\ref{conclusions}.
Appendix~\ref{app:pulls_scbII} contains the list of observables that
contribute to the global fit, as well as their prediction in
the most general scenario considered in this work.

\section{Brief summary of the Effective Field Theory}\label{sec:EFT}

This section presents a short summary of the Effective Field Theory used in our analysis.
First, at energy scales relevant for flavour
processes it is convenient to work at 
an energy scale below the electroweak (EW) scale, for example $\mu_\mathrm{WET} = m_b$,
with the top quark, Higgs, $W$ and $Z$ bosons being integrated out.
The relevant terms of the Weak Effective Theory (WET)
Lagrangian~\cite{Buras:1998raa,Aebischer:2015fzz,Aebischer:2017gaw,Tanaka:2012nw} for the semileptonic decays of $B$ mesons are:
\begin{align}
  \mathcal{L}_{\text{eff}} = & -\frac{4 G_F}{\sqrt{2}}V_{cb}\sum_{\ell = e, \mu, \tau} (1 + C_{VL}^\ell) \mathcal{O}_{VL}^\ell
  + \frac{4G_F}{\sqrt{2}}V_{tb}V_{ts}^*\frac{e^2}{16\pi^2}\sum_{\ell=e,\mu} (C_9^\ell \mathcal{O}_9^\ell  + C_{10}^\ell \mathcal{O}_{10}^\ell) \nonumber          \\
                             & + \frac{4G_F}{\sqrt{2}}V_{tb}V_{ts}^*\frac{e^2}{16\pi^2}\sum_{\ell=e,\mu,\tau} C_\nu^\ell \mathcal{O}_\nu^\ell\ ,\label{eq:lagWET}
\end{align}
where $G_F$ is the Fermi constant, $e$ is the electromagnetic coupling and $V_{qq'}$
are the elements of the Cabibbo-Kobayashi-Maskawa (CKM) matrix. The dimension six
operators are defined as,  
\begin{align}
  \mathcal{O}_{VL}^\ell = (\bar{c}_L \gamma_\alpha b_L)(\bar{\ell}_L \gamma^\alpha \nu_\ell)\ ,     & \qquad \mathcal{O}_9^\ell = (\bar{s}_L \gamma_\alpha b_L)(\bar{\ell} \gamma^\alpha \ell)\ ,\nonumber              \\
  \mathcal{O}_{10}^\ell = (\bar{s}_L \gamma_\alpha b_L)(\bar{\ell} \gamma^\alpha \gamma_5 \ell) \ , & \qquad \mathcal{O}_\nu^\ell = (\bar{s}_L\gamma_\alpha b_L)[\bar{\nu}_\ell \gamma^\alpha (1-\gamma_5)\nu_\ell] \ ,
\end{align}
being their corresponding Wilson coefficients $C_{VL}^\ell$, $C_9^\ell$,
$C_{10}^\ell$ and $C_\nu^\ell$, respectively. The last three coefficients
have contributions from both the SM processes ($C_i^{\mathrm{SM}\,
  \ell}$), and NP contribution ($C_i^{\mathrm{NP}\, \ell}$),
\begin{equation}
C_i^\ell = C_i^{\mathrm{SM}\, \ell} + C_i^{\mathrm{NP}\, \ell}\ ,\qquad\qquad i= 9,10,\nu\ .
\end{equation}

The dependence of the $\RKp$ ratios on the Wilson coefficients has been
previously obtained in~\cite{Alda:2018mfy},
\begin{equation}
R_{K^*}^{[1.1,6]} \simeq \frac{0.9875+0.1759\, \mathrm{Re}\,\CnmuNP -
  0.2954\,  \mathrm{Re}\, \CtmuNP + 0.0212|\CnmuNP|^2 + 0.0350
  |\CtmuNP|^2}{1\,\  \ \ \ \ +0.1760\, \mathrm{Re}\,\CneNP - 0.3013\,
  \mathrm{Re}\,  \CteNP + 0.0212|\CneNP|^2 + 0.0357 |\CteNP|^2}\ .
\label{eq:JA_Rk}
\end{equation}
where an analytic 
computation of this ratio as a function of $\CnmuNP$, $\CtmuNP$ in the
region $1.1 \leq q^2 \leq 6.0\GeV^2$ was performed. 

For the \RDp ratios, the dependence of the \RDp ratios on the Wilson coefficients is given 
by~\cite{Bhattacharya:2016mcc,Feruglio:2017rjo}:
\begin{align}
\RDp^\ell &= \RDp^{\ell, \mathrm{SM}} \frac{|1+ C_{VL}^\tau|^2}{ (|1+C_{VL}^e|^2 + |1+C_{VL}^\mu|^2)/2}\ , \nonumber\\
\RDp^\mu &= \RDp^{\mu, \mathrm{SM}} \frac{|1+ C_{VL}^\tau|^2}{ |1+C_{VL}^\mu|^2}\ . \label{eq:RD}
\end{align}

Second, the NP contributions at an energy scale $\Lambda$ ($\Lambda \sim
\mathcal{O}(\mathrm{TeV})$) is defined via the Standard Model
Effective Field Theory (SMEFT) Lagrangian~\cite{Grzadkowski:2010es},
\begin{equation}
  \mathcal{L}_\mathrm{SMEFT} =
  \frac{1}{\Lambda^2}\left(\Clqo^{ijkl}\, O_{\ell q(1)}^{ijkl} +
    \Clqt^{ijkl}\,  O^{ijkl}_{\ell q(3)}   \right) \ ,
\label{eq:Lagr_SMEFT}
\end{equation}
where $\ell$ and $q$ are the lepton and quark $SU(2)_L$ doublets in the basis of electroweak eigenstates,
and ${i,j,k,l}$ denote generation indices. The dimension six operators are defined as 
\begin{equation}
O_{\ell q(1)}^{ijkl} = (\bar{\ell}'_i \gamma_\mu \ell'_j)(\bar{q}'_k \gamma^\mu  q'_l),\qquad\qquad O_{\ell q(3)}^{ijkl}= (\bar{\ell}'_i \gamma_\mu \tau^I \ell'_j)(\bar{q}'_k \gamma^\mu \tau^I q'_l)
\end{equation}
with $\tau^I$ being the Pauli matrices.

We note that we  will use  the SMEFT operators for our numerical
analysis, and will refer to the WET operators only for discussion and
comparison with other previous results in the literature.

The translation between the SMEFT Lagrangian in the electroweak basis and in
the mass basis was obtained in~\cite{Aebischer:2015fzz}. The SMEFT Lagrangian
in the mass basis is
\begin{align}
  \mathcal{L}_\mathrm{SMEFT}^\mathrm{mass} = & \frac{\widetilde{C}_{\ell q(1)}^{ijkl}}{\Lambda^2}
  \Big(\bar{\nu}_{i\,L}\gamma_\mu \nu_{j\,L} + \bar{e}_{i\,L}\gamma_\mu e_{j\,L}\Big)
  \Big(V_{mk}V^*_{nl}\bar{u}_{m\,L}\gamma^\mu u_{n\,L}+\bar{d}_{k\,L}\gamma^\mu d_{l\,L}\Big) \nonumber \\
                              & +\frac{\widetilde{C}_{\ell q(3)}^{ijkl}}{\Lambda^2}
  \Big(\bar{\nu}_{i\,L}\gamma_\mu \nu_{j\,L} - \bar{e}_{i\,L}\gamma_\mu e_{j\,L}\Big)
  \Big(V_{mk}V^*_{nl}\bar{u}_{m\,L}\gamma^\mu u_{n\,L}-\bar{d}_{k\,L}\gamma^\mu d_{l\,L}\Big) \nonumber \\
                              & +2 \frac{\widetilde{C}_{\ell q(3)}^{ijkl}}{\Lambda^2}
  \Big[(\bar{\nu}_{i\,L}\gamma_\mu e_{j\,L})(V_{mk}\bar{u}_{m\,L}\gamma^\mu d_{l\,L}) + (\bar{e}_{i\,L}\gamma_\mu \nu_{j\,L})(V^*_{nl}\bar{d}_{k\,L}\gamma^\mu d_{n\,L})\Big]\,. \label{eq:SMEFTmass}
\end{align}

The relation between the $\Clq$ coefficients in the electroweak basis
and the $\widetilde{C}_{\ell q}$ coefficients in the mass basis
is given by~\cite{Aebischer:2015fzz}
\begin{equation}
\widetilde{C}_{\ell q(1)}^{ijkl} = C_{\ell q(1)}^{ijmn} (U_{d\,L}^*)_{km} (U_{d\,L})_{ln}\,,\qquad\qquad \widetilde{C}_{\ell q(3)}^{ijkl} = C_{\ell q(3)}^{ijmn} (U_{d\,L}^*)_{km} (U_{d\,L})_{ln}\,,\label{eq:eqbases}
\end{equation}
where $U_{d\,L}$ and $U_{u\,L}$ are the SM rotation matrices
for the left-handed quarks. The only constraint for these matrices
is given by the CKM matrix, $V = U_{u\,L}U^\dagger_{d\,L} $. The choice
$U_{d\,L} = 1$, $U_{u\,L} = V$ defines the ``Warsaw-down'' basis of
the SMEFT~\cite{Aebischer:2017ugx},
where $\Clqo = \widetilde{C}_{\ell q(1)}$ and $\Clqt = \widetilde{C}_{\ell q(3)}$.

Finally, there is a recent proposal that links the $B$ meson anomalies with NP in the top sector \cite{Feruglio:2017rjo,Feruglio:2018jnu,Cornella:2018tfd}. In the interaction basis, denoted by double-primed fermions, only the third generation particles exhibit NP couplings,
\begin{equation}
\mathcal{L}_\mathrm{NP} = \frac{1}{\Lambda^2}[C_1(\bar{\ell}''_3 \gamma_\mu \ell''_3)(\bar{q}''_3 \gamma^\mu q''_3) + C_3(\bar{\ell}''_3 \gamma_\mu \tau^I \ell''_3)(\bar{q}''_3 \gamma^\mu \tau^I q''_3)]\ ,
\label{eq:LagParide}
\end{equation}
where $C_1 = \Clqo^{3333}$ and $C_3 = \Clqt^{3333}$. The interaction
basis is related to the basis where the mass matrices are diagonal via
the unitary transformations, 
\begin{align}
u_L = \hat{U}_u u''_L\,, \qquad\qquad d_L = \hat{U}_d d''_L\,, \qquad\qquad \nu_L = \hat{U}_\ell \nu''_L\,, \qquad\qquad e_L = \hat{U}_\ell e''_L \ ,
\end{align}
where $\psi_L = P_L \psi$ ($\psi = u,\ d,\ \nu,\ e$),
$\hat{U}_\psi$ are unitary matrices, and the quark
unitary matrices are related to the CKM matrix as $\hat{U}_u
  \hat{U}^\dagger_d = V$. The fermionic bilinears are transformed as follows,
\begin{align}
\bar{u}''_3 \gamma_\mu u''_3 = \lambda^u_{ij} \bar{u}_i \gamma_\mu u_j\ ,
\qquad\qquad \bar{d}''_3 \gamma_\mu d''_3 &= \lambda^q_{ij} \bar{d}_i
\gamma_\mu d_j\ , \qquad\qquad \bar{u}''_3 \gamma_\mu d''_3 = \lambda^{ud}_{ij} \bar{u}_i \gamma_\mu d_j \nonumber \\
\bar{e}''_3 \gamma_\mu e''_3 = \lambda^\ell_{ij} \bar{e}_i \gamma_\mu e_j\
,\qquad\qquad \bar{\nu}''_3 \gamma_\mu \nu''_3 &= \lambda^\ell_{ij}
\bar{\nu}_i \gamma_\mu \nu_j\ , \qquad\qquad \bar{e}''_3 \gamma_\mu \nu''_3 = \lambda^\ell_{ij} \bar{e}_i \gamma_\mu \nu_j\ ,
\end{align}
with the flavour matrices $\lambda$ given by
\begin{align}
\lambda^u_{ij} = (\hat{U}_u)_{3i} (\hat{U}_u)^*_{3j}\ , &\qquad \lambda^q_{ij} =
(\hat{U}_d)_{3i} (\hat{U}_d)^*_{3j}\ , \nonumber\\
\lambda^{ud}_{ij} = (\hat{U}_u)_{3i} (\hat{U}_d)^*_{3j}\ , &\qquad
\lambda^\ell_{ij} = (\hat{U}_\ell)_{3i} (\hat{U}_\ell)^*_{3j}\ .
\label{eq:def_lambdamatrices}
\end{align}

We can write all the quark matrices in terms of $\lambda^q$, 
\begin{equation}
\lambda^u = V \lambda^q V^\dagger \,,\qquad\qquad \lambda^{ud} = V \lambda^q\  ,
\end{equation}
so every $u$-type quark picks an additional CKM matrix, which are
exactly the same factors appearing in the Lagrangian for the mass basis
in Eq.~\eqref{eq:SMEFTmass}. For example, if we expand the first term in
Eq.~\eqref{eq:LagParide}, we obtain
\begin{align}
    & \frac{C_1}{\Lambda^2}(\bar{\ell}''_3\gamma_\mu \ell''_3)(\bar{q}''_3\gamma^\mu q''_3) \nonumber                                                                              \\
  = & \frac{C_1}{\Lambda^2}\Big(\bar{\nu}''_3\gamma_\mu \nu''_3 + \bar{e}''_3\gamma_\mu e''_3\Big)\Big(\bar{u}''_3 \gamma^\mu u''_3 + \bar{d}''_3 \gamma^\mu d''_3 \Big) \nonumber \\
  = & \frac{C_1}{\Lambda^2}\lambda^\ell_{ij} \lambda^q_{kl} \Big(\bar{\nu}_{i\,L}\gamma_\mu \nu_{j\,L} + \bar{e}_{i\,L}\gamma_\mu e_{j\,L}\Big)
  \Big(V_{mk}V^*_{nl}\bar{u}_{m\,L}\gamma^\mu u_{n\,L}+\bar{d}_{k\,L}\gamma^\mu d_{l\,L}\Big)\,,
\end{align}
which agrees with Eq.~\eqref{eq:SMEFTmass} with the identification
$\Clqo^{ijkl} = \widetilde{C}_{\ell q(1)}^{ijkl} = C_1 \lambda^\ell_{ij}
  \lambda^q_{kl}$. Repeating the same steps with the other term in
Eq.~\eqref{eq:LagParide}, we arrive to $\Clqt^{ijkl} =
  \widetilde{C}_{\ell q(3)}^{ijkl} = C_3 \lambda^\ell_{ij}\lambda^q_{kl}$.

In conclusion, the Lagrangian of Eq.~\eqref{eq:LagParide} in the
``Warsaw-down'' basis becomes
\begin{equation}
\mathcal{L}_\mathrm{NP} = \frac{\lambda^\ell_{ij} \lambda^q_{kl}}{\Lambda^2}
\left(C_1(\bar{\ell}_i \gamma_\mu  \ell_j)(\bar{q}_k \gamma^\mu 
  q_l)  + C_3 (\bar{\ell}_i \gamma_\mu \tau^I \ell_j)(\bar{q}_k
  \gamma^\mu \tau^I q_l)\right)\,. 
\label{eq:LagLambda}
\end{equation}

We perform the Renormalization Group (RG) running of the SMEFT Wilson
coefficients from $\Lambda = 1\,\mathrm{TeV}$ down to $\mu_\mathrm{EW}$
\cite{Celis:2017hod}, where we match the SMEFT and WET operators
\cite{Jenkins:2017jig}, and finally we perform the RG running of the WET
coefficients down to $\mu = m_b$ \cite{Jenkins:2017dyc}. We check that
the analytical expressions are in agreement with the numerical results
obtained by the package \texttt{Wilson} \cite{Aebischer:2018bkb}.
This operation is performed for all the effective operators in the WET
that receive contributions from the Lagrangian in Eq.~\eqref{eq:LagLambda}.
Here we reproduce the matching conditions for the Wilson coefficients with
the largest impact on the semileptonic $B$ meson decays, that is,
$C_9^{i\,\mathrm{NP}}$ and $C_{10}^{i\,\mathrm{NP}}$ for the
$B\to K^{(*)}\ell^+\ell^-$ decays, $C_{VL}^{i\,\mathrm{NP}}$ for the
$B \to D^{(*)}\ell\nu$ decays, and $C_\nu^i$ for the $B\to
K^{(*)}\nu\bar{\nu}$
decays:
\begin{align}
C_9^{\mathrm{NP}\,i} &\approx \frac{2\sqrt{2}\pi^2}{e^2 V_{tb} V_{ts}^*}
\frac{1}{G_F
  \Lambda^2}(C_1+C_3)\lambda^q_{23}\lambda^\ell_{ii}
+ \frac{\sqrt{2}}{3 V_{tb}V_{ts}^* } \frac{1}{G_F
  \Lambda^2}(C_1+C_3) \lambda^q_{23}
\log\frac{m_b}{\Lambda}\,, \nonumber\\
C_{10}^{\mathrm{NP}\,i} &\approx -\frac{2\sqrt{2}\pi^2}{e^2 V_{tb}
  V_{ts}^*} \frac{1}{G_F
  \Lambda^2}(C_1+C_3)\lambda^q_{23}\lambda^\ell_{ii}\,,
\nonumber\\
C_{VL}^i &\approx -\frac{1}{\sqrt{2} G_F
           \Lambda^2}C_3\lambda^\ell_{ii}\left(
           \frac{V_{cs}}{V_{cb}}\lambda^q_{23} +
\lambda^q_{33}\right)\,,\nonumber\\ 
C_\nu^i &\approx \frac{2\sqrt{2}\pi^2}{e^2 V_{tb}V_{ts}^*}\frac{1}{G_F \Lambda^2}(C_1-C_3)\lambda^q_{23}\lambda^\ell_{ii} + \frac{3\sqrt{2} {g'}^2}{2 e^2 V_{tb}V_{ts}^*}\frac{1}{G_F \Lambda^2} C_3 \lambda^q_{23}\lambda^\ell_{ii}\log\frac{m_b}{\Lambda}\,.
           \label{eq:matching}
\end{align}

We find out that there is a sizeable subleading term that affects
$C_9^{\mathrm{NP}\,i}$ and not $C_{10}^{\mathrm{NP}\,i}$, thus breaking
the leading-order relation $C_9^{\mathrm{NP}\,i} = -
C_{10}^{\mathrm{NP}\,i}$. However, this subleading term is LFU, since it
does not depend on the leptonic rotation matrix $\lambda^\ell$, and
consequently has a negligible effect on the universality ratios
$\RKp$. The spoiling of the tree-level relation will become relevant in
observables that include only one lepton flavour, such as the branching
ratios and angular observables for $B\to K^{(*)}\mu^+ \mu^-$; which depend on
$C_9^\mu$ and $C_{10}^\mu$; and $B_s \to \mu^+ \mu^-$ only depending on
$C_{10}^\mu$. The interplay between the tree-level and loop-induced
terms is well known and was also previously discussed
by~\cite{Cornella:2021sby}. 

In order to describe the rotation from the two bases, the $\lambda$ matrices
introduced in Eq.~\eqref{eq:def_lambdamatrices} must be hermitian,
idempotent $\lambda^2 = \lambda$, and $\mathrm{tr}\lambda = 1$.
These properties are consequences of the fact that, in the interaction basis,
NP only affects one generation, and follow immediately from the definitions:
\begin{align}
  \lambda_{ji} & = U^*_{3j} U_{3i} = (U^*_{3i} U_{3j})^* =
                 \lambda^*_{ij}\,,\nonumber \\
  \lambda_{ij}\lambda_{jk} & = U^*_{3i} U_{3j} U^*_{3j} U_{3k} =
                             U^*_{3i}U_{3k} = \lambda_{ik}\,,\nonumber \\
  \mathrm{tr}\lambda & = \sum_i \lambda_{ii} = \sum_i U^*_{3i} U_{3i} = (U^\dagger U)_{33} = 1\,.
\end{align}
A $3\times 3$ hermitian idempotent matrix with trace one has 4 free real
parameters, or equivalently, 2 free complex parameters. Without loss of
generality, we can use the parameterization~\cite{Feruglio:2017rjo}
\begin{equation}
\lambda^{\ell,\,q} = \frac{1}{1 + |\alpha^{\ell,\,q}|^2  +
  |\beta^{\ell,\,q}|^2}
\begin{pmatrix}
|\alpha^{\ell,\,q}|^2 & \alpha^{\ell,\,q}\bar{\beta}^{\ell,\,q} & \alpha^{\ell,\,q} \\ \bar{\alpha}^{\ell,\,q}\beta^{\ell,\,q} & |\beta^{\ell,\,q}|^2  & \beta^{\ell,\,q} \\ \bar{\alpha}^{\ell,\,q} & \bar{\beta}^{\ell,\,q} & 1
\end{pmatrix}\,, \label{eq:Idemp}
\end{equation}
where $\alpha^{\ell,\,q}$ and $\beta^{\ell,\,q}$ are complex numbers, which
are related to the unitary rotation matrices as
\begin{align}
  (U_{\ell,q})_{31} = \frac{\bar{\alpha}^{\ell,q}}{\sqrt{1+|\alpha^{\ell,q}|^2 +
  |\beta^{\ell,q}|^2}}\,,\nonumber \\
  (U_{\ell,q})_{32} = \frac{\bar{\beta}^{\ell,q}}{\sqrt{1+|\alpha^{\ell,q}|^2 +
  |\beta^{\ell,q}|^2}}\,,\nonumber  \\
  (U_{\ell,q})_{33} = \frac{1}{\sqrt{1+|\alpha^{\ell,q}|^2 + |\beta^{\ell,q}|^2}}\,,\nonumber                     \\
  \alpha^{\ell,q} = \frac{(U_{\ell,q}^*)_{31}}{(U_{\ell,q})_{33}}\,\qquad
  \beta^{\ell,q} = \frac{(U_{\ell,q}^*)_{32}}{(U_{\ell,q})_{33}}\,
\end{align}

We can therefore understand the parameters $\alpha^\ell$ and $\beta^\ell$ as the
relative degree of mixing to the first and second generations of leptons,
respectively, produced by the rotation from the interaction basis to the mass
basis. Analogously, the parameters $\alpha^q$ and $\beta^q$ represent the
relative degree of mixing to the first and second generations of $d$-type quarks
(remember that the $u$-type quarks pick additional CKM factors).

The conditions in Eq.~\eqref{eq:Idemp} impose several relations between the
LFUV operators, which are proportional to the diagonal entries of
$\lambda^\ell$, and the LFV operators, proportional to the off-diagonal entries:
\begin{equation}
\Clq^{11ij} = \frac{|\Clq^{13ij}|^2}{\Clq^{33ij}}\qquad\qquad \Clq^{22ij} = \frac{|\Clq^{23ij}|^2}{\Clq^{33ij}}\,.
\end{equation}

On the other hand, the $\mathcal{O}_{\ell q}$ operators also produce unwanted contributions
to the $B \to K^{(*)} \nu \bar{\nu}$ decays~\cite{Feruglio:2017rjo}. In
order to obey these constraints, we will fix at the scale $\mu=\Lambda$
the relation 
\begin{equation}
  \Clqo^{ijkl} = \Clqt^{ijkl} \equiv \Clq^{ijkl}\ .
  \label{eqc1c3}
\end{equation}
This relation cancels the tree-level contribution to the $B \to K^{(*)}
\nu \bar{\nu}$, but there is still a loop-induced contribution,
proportional to the $\Clqt$ coefficients. However, we have checked that
in our scenarios, this is only a 0.1\% correction of the SM predictions.

\section{Global fits}\label{sec:Fits}

Since the effective operators affect a large number of observables, connected
between them via the Wilson coefficients, any NP prediction based on
Wilson coefficients has to be confronted not
only with the \RKp an \RDp measurements, but also with additional
several measurements involving the  $B$-mesons decays.
The RG evolution in the SMEFT produces a mix of the
low-energy effective operators, which implies that
the $W$ and $Z$ couplings to leptons are
modified~\cite{Jenkins:2013wua,Alonso:2013hga}.
Then, several EW observables are affected, such as the $W$ boson mass,
the hadronic cross-section of the $Z$ boson 
($\sigma^0_\mathrm{had}$) or the branching ratios of the $Z$ to different
leptons. In order to keep the predictions consistent with this range of
experimental test, global fits have proven to be a valuable
tool~\cite{Celis:2017doq,Camargo-Molina:2018cwu,Aebischer:2019mlg,Aoude:2020dwv}.       
We have previously done in~\cite{Alda:2020okk, Alda:2021ruz} an analysis of the
effects of the global fits to the Wilson
coefficients, assuming a model independent effective Hamiltonian
approach. 

In the current paper the global fits to the $\Clq$ Wilson
coefficients have been performed by using the packages
\texttt{flavio} v2.3~\cite{Straub:2018kue} and
\texttt{smelli} v2.3~\cite{Aebischer:2018iyb}\footnote{We have supplemented
  the experimental measurements of the \texttt{smelli} v2.3 database with
  additional $B\to K^* e^+ e^-$~\cite{LHCb:2015ycz} and
  $B\to K^* \mu^+ \mu^-$~\cite{LHCb:2020lmf} angular observables.}.
This code assumes unitarity of the CKM matrix.
Note that the experimental measurements used to determine the SM
input parameters, such as the $\mu\to e\bar{\nu}\nu$ decay, are not
included in the fit in order to ensure the consistency of the
procedure.

In our analysis, the goodness of each fit is evaluated with its
difference of $\chi^2$ with respect to the SM,
$\Delta \chi^2_\mathrm{SM} = \chi^2_\mathrm{SM} -
\chi^2_\mathrm{fit}$. The package \texttt{smelli} actually computes the
differences of the logarithms of the likelihood function $\Delta \log L
= -\frac{1}{2} \Delta \chi^2_\mathrm{SM}$. In order to compare two 
fits $A$ and $B$, we use the pull between them in units of $\sigma$, 
defined as~\cite{Descotes-Genon:2015uva,Capdevila:2018jhy} 
\begin{equation}
\mathrm{Pull}_{A \to B} = \sqrt{2} \mathrm{Erf}^{-1}[F(\Delta \chi^2_A - \Delta \chi^2_B; n_B - n_A )]\,,
\end{equation}
where $\mathrm{Erf}^{-1}$ is the inverse of the error function, $F$ is
the cumulative distribution function of the $\chi^2$ distribution and
$n$ is the number of degrees of freedom of each fit.\\
The SM input parameters used for these fits are the same as in our
previous work~\cite{Alda:2020okk}. The Renormalization Group effects of
the SMEFT operators that shift the  Fermi constant
$G_F$~\cite{Jenkins:2017jig} from its SM value $G_F^0$ are considered.
The effects on the CKM matrix~\cite{Descotes-Genon:2018foz} are not
implemented, and its parameters are treated as nuance parameters
instead.

Now we proceed to fit the set of flavour observables to the parameters
$C_1=C_3 \equiv C$, $\alpha^{\ell,\,q}$ and $\beta^{\ell,\,q}$ of
Eqs.~\eqref{eq:LagLambda} and~\eqref{eq:Idemp}. In this setting, we
consider two Scenarios: 
\begin{itemize}
\item \textbf{Scenario I:} $\lambda_{11}^{\ell,\,q} =
  \lambda_{12}^{\ell,\,q} = \lambda_{13}^{\ell,\,q} = 0$,
  that is, $\alpha^\ell = \alpha^q = 0$, and $C_1 = C_3$.
\item \textbf{Scenario II:} The only assumption is $C_1 = C_3$.
\end{itemize}

In both scenarios $C_1 = C_3$ in order to implement the constraints
from the $B\to K^{(*)}\bar{\nu}\nu$ observables, as previously
mentioned (see Eq.~(\ref{eqc1c3})). In Scenario I we also
set $\lambda_{11}^{\ell,\,q} = \lambda_{12}^{\ell,\,q} =
\lambda_{13}^{\ell,\,q} = 0$, i.e. $\alpha^\ell = \alpha^q = 0$,
assuming that the mixing affecting the first generation are negligible;
this is the same assumption used in~\cite{Feruglio:2017rjo}. Scenario II
is more general, including non-negligible mixings to the first
generation, allowing us to check the validity of the above assumption
and to discuss the results in a more general situation; focusing in
the relevance of the mixing in the first generation. 
In both scenarios, we only consider real values for the parameters of the fit.

\begin{figure}
\centering
\begin{tabular}{cc}
  \includegraphics[width=0.45\textwidth]{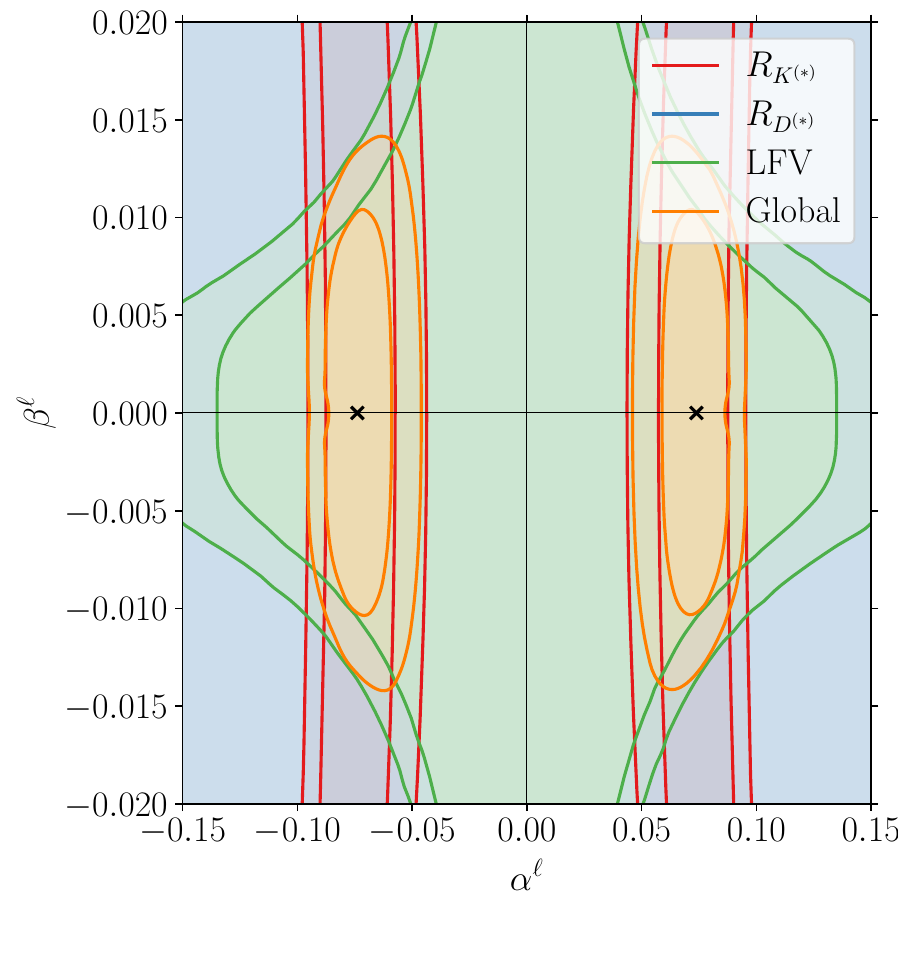}&
  \includegraphics[width=0.45\textwidth]{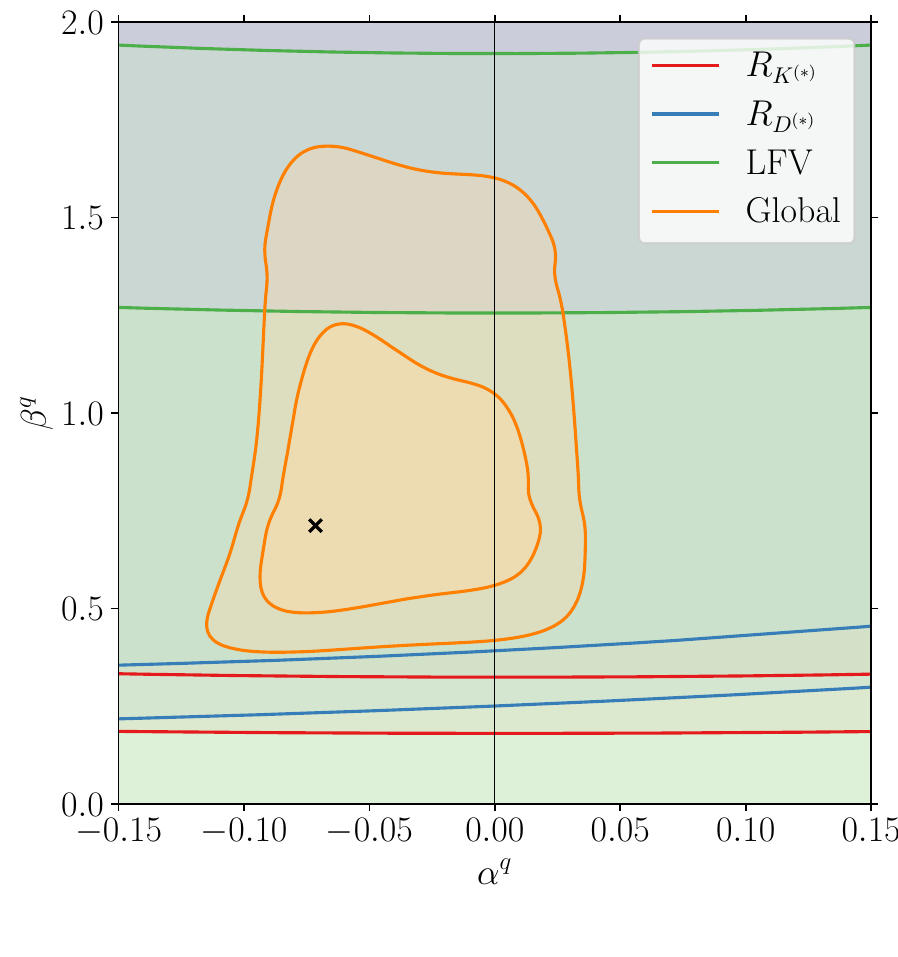}\\
(a) & (b)
\end{tabular}
\caption{$1 \sigma$ and $2 \sigma$ contours for the (a) $\alpha^\ell$
  and $\beta^\ell$ and (b) $\alpha^q$ and $\beta^q$ parameters, with
  the rest of parameters as in the best fit point of Scenario II.}
\label{im:alphabeta}
\end{figure}

\begin{table}
\centering
\begin{tabular}{|c|c|c|}\hline
	&	Scenario I	&	Scenario II	\\\hline
$C$	&	$-0.12 \pm 0.05$	&	$-0.123 \pm 0.010$ \\\hline
$\alpha^\ell$	&		&	$\pm (0.074 \pm 0.024)$	\\\hline
$\beta^\ell$	&	$0 \pm 0.025$	&	$0 \pm 0.019$	\\\hline
$\alpha^q$ & & $-0.07^{+0.08}_{-0.01}$ \\\hline
$\beta^q$ & $0.78^{+1.70}_{-0.4}$ & $0.71^{+1.1}_{-0.47}$ \\\hline
$\Delta \chi^2_\mathrm{SM}$	&	41.37	&	58.84	\\\hline
SM Pull	&	5.83 $\sigma$	&	6.70 $\sigma$	\\\hline
$p$-value & $5.5\times 10^{-9}$ & $2.1\times 10^{-11}$ \\\hline
\end{tabular}
\caption{Best fits to the rotation parameters and the coefficient
  $C$ in Scenarios I and II.}\label{tab:rotation} 
\end{table}
The best fits to the rotation parameters $\alpha$ and $\beta$ for 
leptons and quarks and to the Wilson coefficient $C\equiv C_1=C_3$ 
in these two Scenarios are summarized in
Table~\ref{tab:rotation}. The best fit is found for Scenario II, with
a pull of 6.70 $\sigma$ with respect to the Standard Model, 3.77 $\sigma$
with respect to Scenario I. We note that the $\beta^\ell$ parameter,
which mixes the second and third generations of leptons at tree level, is
negligible in both fits.
Figure~\ref{im:alphabeta} shows the two-dimensional sections of the
likelihood function $\Delta \chi^2_\mathrm{SM}$ for the $\alpha^\ell$-$\beta^\ell$
and $\alpha^q$-$\beta^q$  parameters in Scenario II, at $1 \sigma$ and
$2 \sigma$. The rest of parameters are given as in the best fit point
of this Scenario. Results for the $\RKp$ and $\RDp$ observables and
for the LFV observables, as well as for the global fit are included. 
We can observe that, due to the
non-linear relations imposed by Eq.~\eqref{eq:Idemp}, the regions of
equal probability are highly non-ellipsoidal. Therefore, we cannot use
the Gaussian approximation to characterise the fit. Instead, we will use
a Montecarlo analysis, described in section \ref{sec:MC}, in order to
extract the confidence intervals and correlations between observables. The
values of the parameters of the Lagrangian~(\ref{eq:LagLambda}) in
Scenario II are $C = C_1 = C_3= -0.126 \pm 0.010$, and 
\begin{equation}
\lambda^\ell =  \begin{pmatrix}
(5 \pm 4) \times 10^{-3} &  (0 \pm 9) \times 10^{-4} & (7 \pm 3) \times 10^{-2} \\
(0 \pm 9) \times 10^{-4} & (0\pm 2)\times 10^{-4} & (0 \pm 2) \times 10^{-2} \\
(7 \pm 3) \times 10^{-2} & (0 \pm 2) \times 10^{-2} &  0.995 \pm 0.004
\end{pmatrix}\ ,
\end{equation}

\begin{equation}
\lambda^q = \begin{pmatrix}
(3^{+0}_{-3}) \times 10^{-3} & (-3^{+2}_{-0}) \times 10^{-2} & (-4^{+2}_{-0}) \times 10^{-2} \\
(-3^{+2}_{-0}) \times 10^{-2} & 0.34 \pm 0.29 & 0.47\pm 0.09 \\
(-4^{+2}_{-0}) \times 10^{-2} &  0.47 \pm 0.09 & 0.66 \pm 0.29
\end{pmatrix}\ .
\end{equation}

The most notable effect of the mass rotation is the mixing of the second
and third generation quarks, and there is also some mixing between the
first and third generation leptons.

The more relevant WET Wilson coefficients in Scenario II are 
\begin{align}
C_9^{\mathrm{NP}\,\mu} &= -0.6\pm0.2, \qquad
&&C_{10}^{\mathrm{NP}\,\mu} = -0.002\pm0.01, \qquad C_{VL}^{\mathrm{NP}\,\tau}
= 0.09\pm0.03\,, \nonumber \\
C_9^{\mathrm{NP}\,e} &= -0.25\pm0.21, \qquad\qquad
&&C_{10}^{\mathrm{NP}\,e} = -0.36\pm0.23.
\end{align}

As established in Eq.~(\ref{eq:matching}), subleading RG effects cause a
notable deviation from the leading-order relation $C_9^{\mathrm{NP}\,\mu} = -
C_{10}^{\mathrm{NP}\,\mu}$. This is in agreement with the fits performed
in~\cite{Descotes-Genon:2015uva,Hurth:2016fbr,Altmannshofer:2017fio,Capdevila:2017bsm,Altmannshofer:2017yso,DAmico:2017mtc,Geng:2017svp,Ciuchini:2017mik,Alda:2018mfy,Alok:2019ufo,Kumar:2019nfv,Capdevila:2019tsi,Bhattacharya:2019dot,Biswas:2020uaq,Bhom:2020lmk},
where the Wilson coefficient $C_9^{\mathrm{NP}\,\mu}$ receives a greater
NP contribution than $C_{10}^{\mathrm{NP}\,\mu}$. According to our fit,
$C_{10}^{\mathrm{NP}\,\mu} \approx 0$: from the matching conditions,
this operator is generated at tree level and is proportional to
$\lambda^\ell_{22} ~ \sim |\beta^\ell|^2$. From the plot in
Fig.~\ref{im:alphabeta} we learn that the parameter $\beta^\ell$ is
severely constrained by the LFV observables, in green lines.
Consequently $C_9^{\mathrm{NP}\,\mu} = - C_{10}^{\mathrm{NP}\,\mu} +
C_9^\mathrm{loop} \approx C_9^\mathrm{loop}$ is dominated by the
loop-generated term in Eq.~\eqref{eq:matching}. Clearly, the logarithmic
term that appear in the first equation of~\eqref{eq:matching} is
relevant in the phenomenological analysis. In the electron sector,
the mixing parameter $\alpha^\ell$ does not suffer large constraints
from the LFV sector. In this case,
the tree-level and loop-level terms are similar, and therefore
$C_9^{\mathrm{NP}\,e} = - C_{10}^{\mathrm{NP}\,e} + C_9^\mathrm{loop}
\approx - C_{10}^{\mathrm{NP}\,e} +C_9^{\mathrm{NP}\,\mu}$, which is of
the same order of magnitude as $C_{10}^{\mathrm{NP}\,e}$.
In Section~\ref{sec:simplifiedmodel}, we assess an specific
model of leptoquarks where these relations are met.

\begin{figure}
\centering
\begin{tabular}{cc}
  \includegraphics[width=0.45\textwidth]{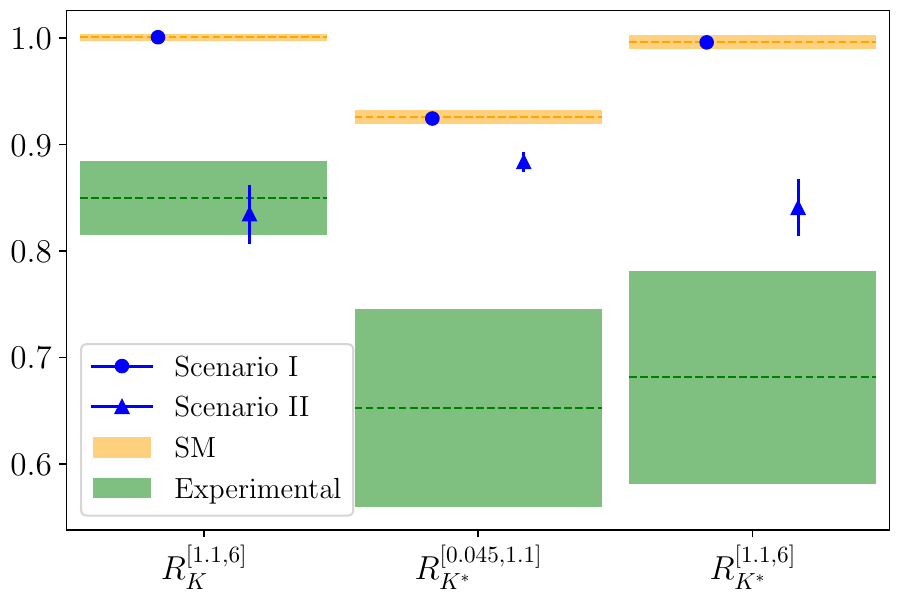}&
  \includegraphics[width=0.45\textwidth]{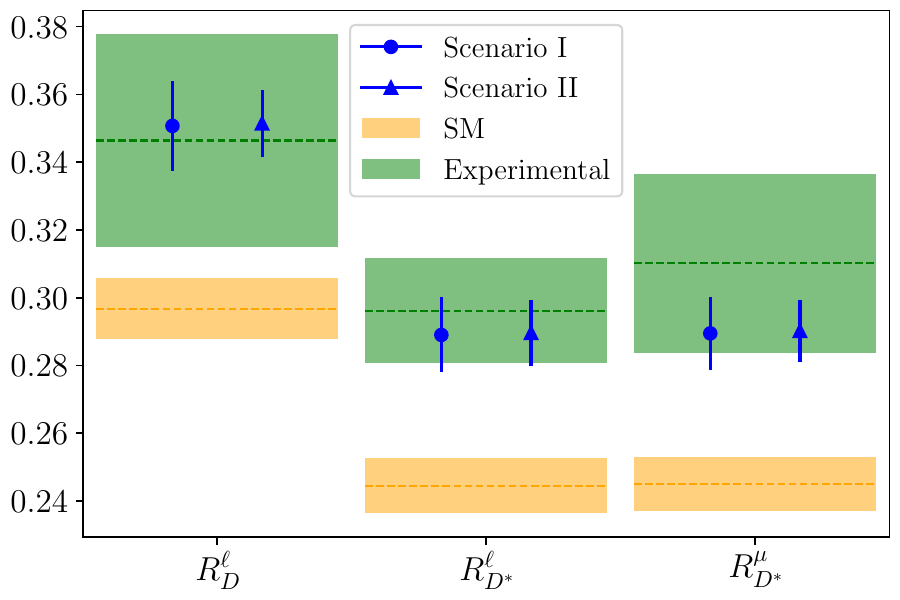}\\
(a)&(b)
\end{tabular}
\caption{Central value and $1 \sigma$ uncertainty (blue lines) of the
  (a) $\RKp$ observables and (b) $\RDp$ observables in Scenarios I and
  II, compared to the Standard Model prediction (yellow area) and
  experimental measurements (green area).} 
\label{im:obs_scbII}
\end{figure}
The predictions for \RKp and \RDp observables in the best fit points
for both scenarios are displayed in Figure~\ref{im:obs_scbII}, where the 
central value and $1 \sigma$ uncertainty of the observables are included.
The yellow area corresponds with the SM prediction, and the green area
with the experimental measurements for each observable.
Table~\ref{tab:observables_rot} summarizes
the results for the $\RKp$ and $\RDp$ observables in Scenarios I and II
for the corresponding best fit points.
For comparison, an statistical combination of all the available
measurements of each observable, performed by \texttt{flavio} is included in the
last column of this table.

From the above results, it is clear that the
assumptions of Scenario I do not allow for a simultaneous explanation of
the \RKp and \RDp anomalies, as already pointed out in
\cite{Feruglio:2017rjo}.
In particular, a value of the mixing between the second
and third generation leptons $\beta^\ell$ is large enough to describe \RKp
through the tree-level $C_9^{\mathrm{NP}\,\mu} = -
C_{10}^{\mathrm{NP}\,\mu}$ coefficients, but implies that $\RDp <
\RDp^\mathrm{SM}$. Instead, our fit shows a preference for a negligible
$\beta^\ell$, and therefore the  \RDp anomalies are explained only
through NP in $C_{VL}^\tau$. The predictions for the branching
ratios and angular observables of the $B\to K^{(*)}\mu^+ \mu^-$ decays
are improved thanks to the flavour-universal loop-induced contribution
to $C_9^\mathrm{NP} = C_9^\mathrm{loop}$, while the \RKp ratios are not
sensible to the universal contribution and remain SM-like. 

\begin{table}
\centering
\begin{tabular}{|c|c|c|c|}\hline
Observable & Scenario I & Scenario II & Measurement \\\hline
$R_K^{[1.1, 6]}$ & $1.0009 \pm 0.0002$ & $0.83\pm0.03$ & $0.85 \pm 0.06$ \\\hline
$R_{K^*}^{[0.045,\ 1.1]}$ & $0.9244 \pm 0.0005$ & $0.884 \pm 0.010$ & $0.65 \pm 0.09$\\\hline
$R_{K^*}^{[1.1,\ 6]}$ & $0.996 \pm 0.002$ & $0.84\pm0.03$  & $0.68 \pm 0.10$ \\\hline
$R_D^\ell$ & $0.351 \pm 0.013$ & $0.351\pm0.010$ & $0.35 \pm 0.03$ \\\hline
$R_{D^*}^\ell$ & $0.289\pm0.011$ & $0.290 \pm 0.010$ & $0.296 \pm 0.016$ \\\hline
$R_{D^*}^\mu$ & $0.289 \pm 0.011$ & $0.290\pm0.009$ & $0.31 \pm 0.03$ \\\hline
\end{tabular}
\caption{Values of the $\RKp$ and $\RDp$ observables in Scenarios I and
  II for the best fit points.}
\label{tab:observables_rot}
\end{table}
\begin{figure}
\centering
\includegraphics[width=0.65\textwidth]{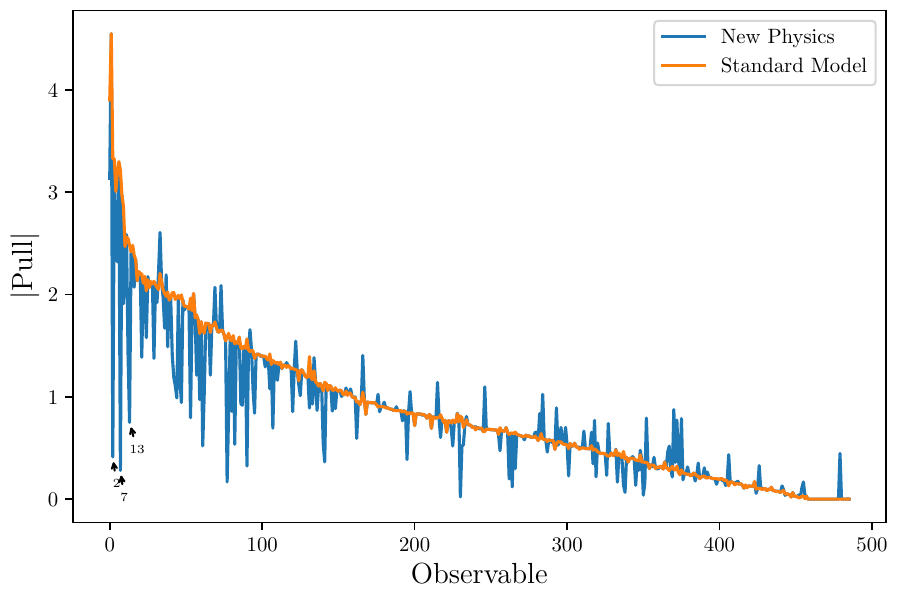}
\caption{Pulls in the Standard Model (orange) and Scenario II (blue) of the observables
  included in the global fit. The observables whose pull changes in more than $1.5\sigma$
  between the SM and Scenario II are specially marked in the plot.} 
\label{im:pulls_scbII}
\end{figure}
The parameters in the fit of Scenario II, on the other hand, are able to describe the
$\RKp$ and $\RDp$ anomalies at the same time, as it is shown in
Figure~\ref{im:obs_scbII} and Table~\ref{tab:observables_rot}. To
consider the mixing between the first and third lepton generation
does not notably alter the prediction for $\RDp$. At the same time, it
originates a tree-level contribution to $C_9^{\mathrm{NP}\,e} =
-C_{10}^{\mathrm{NP}\,e}$, that breaks the universality between the
electron and muon Wilson coefficients, allowing for $\RKp \neq 1$. 
The comparison of the pull of each observable for this scenario
with respect to their experimental measurement (blue line),
compared to the same pull in the SM (orange line) is presented in
Figure~\ref{im:pulls_scbII}. The observables whose pull changes in
more than $1.5\,\sigma$ between the SM and Scenario II are specially
marked in the plot, i.e. $\RDp^l$, $\Rk$  and $\RDp^\mu$
(observables 2, 7 and 13 in the table presented in
Appendix~\ref{app:pulls_scbII}). It is clear that for these observables
NP improves their prediction.
For completeness, the full list of predictions and
pulls is also included in Appendix~\ref{app:pulls_scbII}. We have
checked that all the observables in the appendix, with the only
exception of $|\epsilon_K|$ (observable 33), can receive a contribution
from the Wilson coefficients in Scenario II when considering the full RG
equations. It is also important to note that the muon lifetime is not
included in the above list of observables because it is used to determine the SM value
of $G_F$; an input parameter.

\begin{figure}
  \centering
  \begin{tabular}{ccc}
    \includegraphics[width=0.32\textwidth]{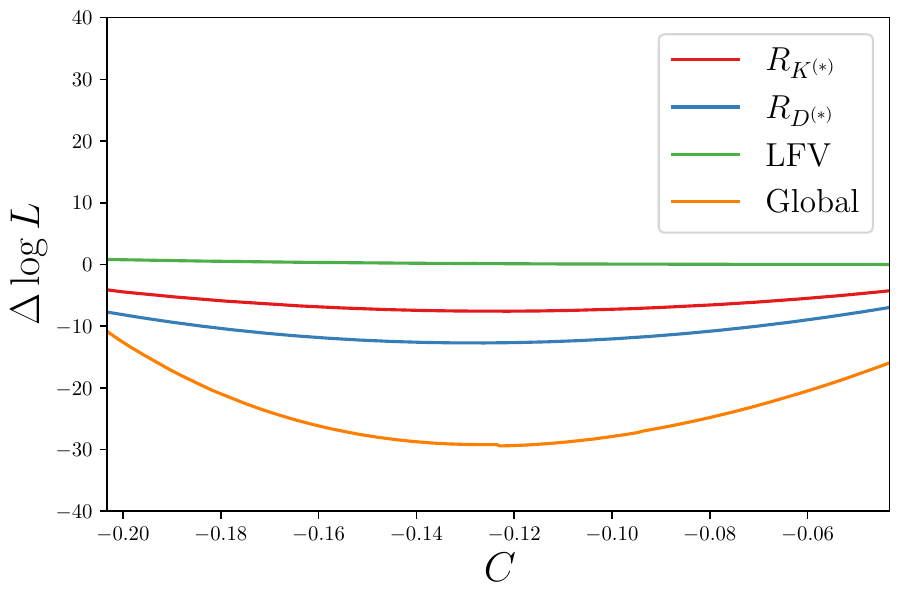} & \includegraphics[width=0.32\textwidth]{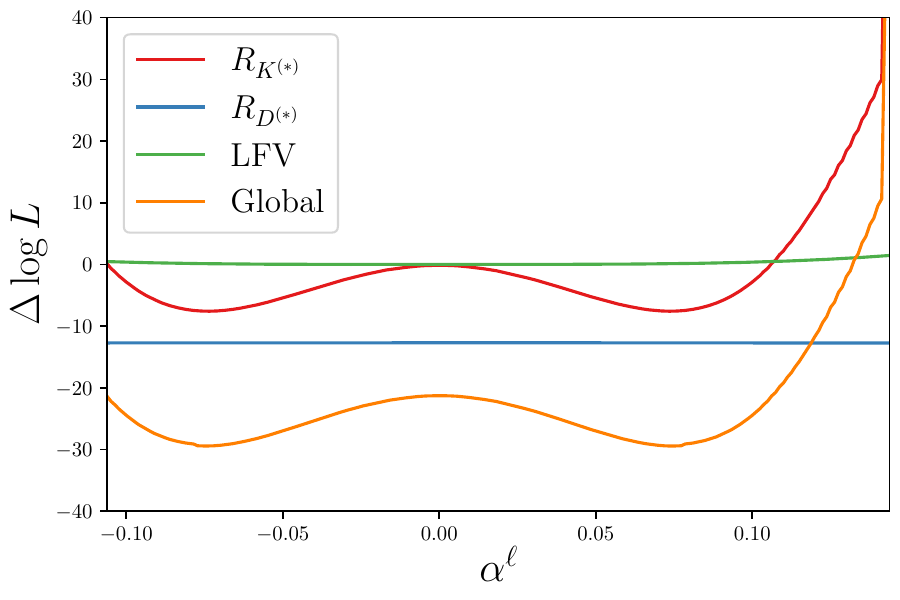} & \includegraphics[width=0.32\textwidth]{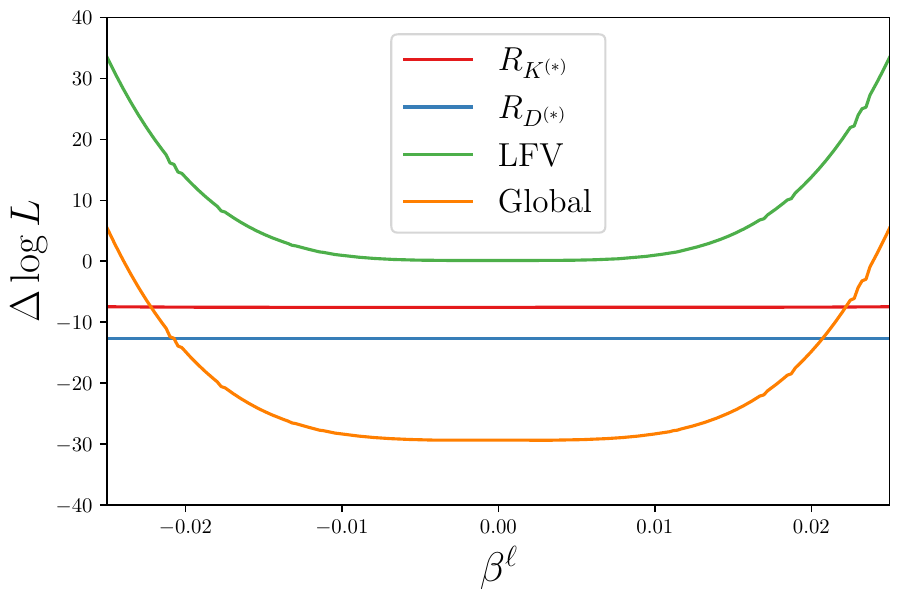} \\
    (a)                                               & (b)                                                & (c)                                                \\
  \end{tabular}
  \begin{tabular}{cc}
    \includegraphics[width=0.32\textwidth]{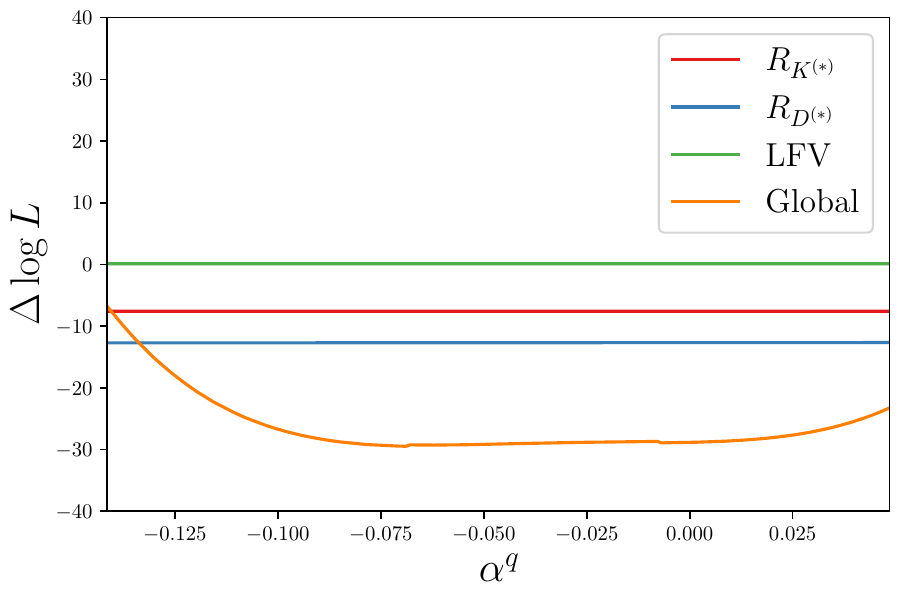} & \includegraphics[width=0.32\textwidth]{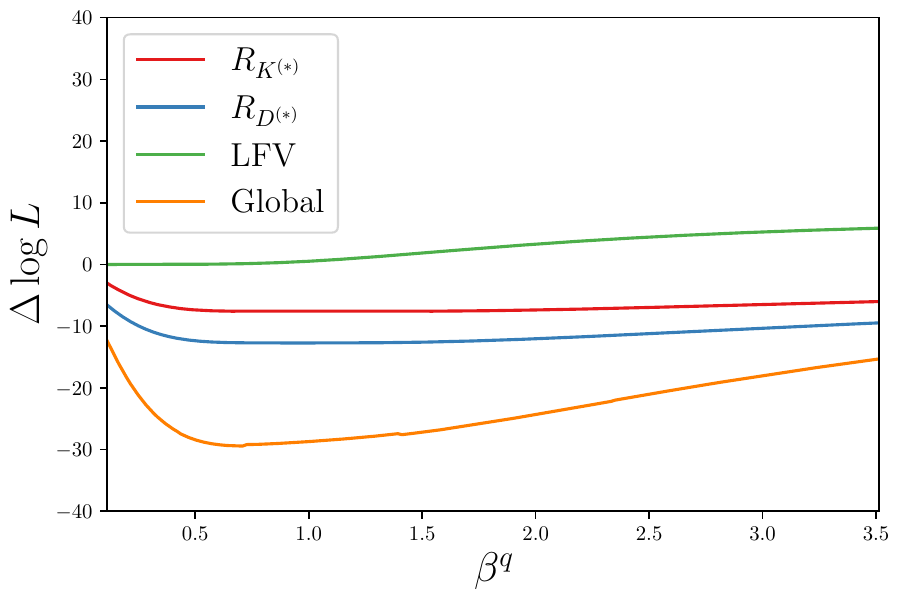} \\
    (d)                                                & (e)
  \end{tabular}
  \caption{Likelihood of the fit when one coefficient is modified: (a) $C$, (b) $\alpha^\ell$, (c) $\beta^\ell$, (d) $\alpha^q$, (e) $\beta^q$}
  \label{im:likelihood_alphabeta}
\end{figure}
Finally, we also investigate which class of observables constraint each
parameter of the fit. For this purpose we modify the rotation parameters
$\alpha$ and $\beta$ for leptons and quarks
and the Wilson coefficient $C\equiv C_1=C_3$ independently, and we compare the results
with respect to the likelihood for $\RKp$, $\RDp$ and LFV observables,
and to the global likelihood.
Figure~\ref{im:likelihood_alphabeta} shows the
evolution of the likelihood for $\RKp$ and $\RDp$ observables and LFV observables, as
well as the global likelihood, when one parameter is modified from its
best fit value. The interplay between all observables is
clearly established when the Wilson coefficient $C$ is modified
(Figure~\ref{im:likelihood_alphabeta} (a)).
In the case of the lepton mixing, it is clear that the
$\RKp$ observables determine the best values of $\alpha^\ell$
(Figure~\ref{im:likelihood_alphabeta} (b)), while the
LFV observables limit the allowed values of $\beta^\ell$ to a narrow
region around zero; being the observables that determine the behaviour
of the global fit in this case (Figure~\ref{im:likelihood_alphabeta} (c)).
In the quark mixing (Figure~\ref{im:likelihood_alphabeta} (d) and (e)),
we found that $\alpha^q$ is constrained by
the observable $\mathrm{BR}(K^+ \to \pi^+ \nu \bar{\nu})$ (observable 406),
while $\beta^q$ is determined by the interplay of $\RKp$ and $\RDp$, that
prefer larger values, and the LFV observables, that disallow $\beta^q > 1$.

Clearly, the above results show the interplay between all parameters and confirm
the relevance of considering all observables when performing phenomenological studies
in the context of $B$-anomalies and the discussion of possible explanation of these anomalies
through NP models.

\section{Montecarlo analysis using Machine Learning}
\label{sec:MC}

In this section we study the parameter points in the neighbourhood of
the best fit point. We will generate samples of parameter points
following the $\chi^2$ distribution given by the likelihood of the
fit. The Montecarlo algorithm is the standard procedure to generate
samples that follow a known distribution. In our case, the computation
time needed to calculate the likelihood of each candidate point is a
huge drawback. Instead, we opted to use a Machine Learning algorithm to
construct an approximation to the likelihood function and that can be
evaluated in a much shorter time. As far as we know, this is the first
time that these procedure is used in the analysis of flavour anomalies.
There exist a previous paper that address the problem of NP model in $b \to c \tau
\nu_\tau$ decays by using a specific machine learning
algorithm~\cite{Bhattacharya:2020vme},
but the techniques used in this paper are different to the ones we used
here.
In the following we give some details of the Machine Learning
procedure and then, we present our results.

\subsection{Methodology}\label{sec:ML_method}

The first Machine Learning tool that we will use for our analysis is a
model able to approximate any arbitrary function $f:\mathbb{R}^n \to
  \mathbb{R}$, that we will use to create an approximation of the log-likelihood
function of our fit. We have chosen an ensemble method based on regression
trees, which is implemented by the XGBoost (eXtreme Gradient Boosting)
algorithm~\cite{Chen:2016}.

\begin{figure}
  \centering
  \includegraphics[width=0.8\textwidth]{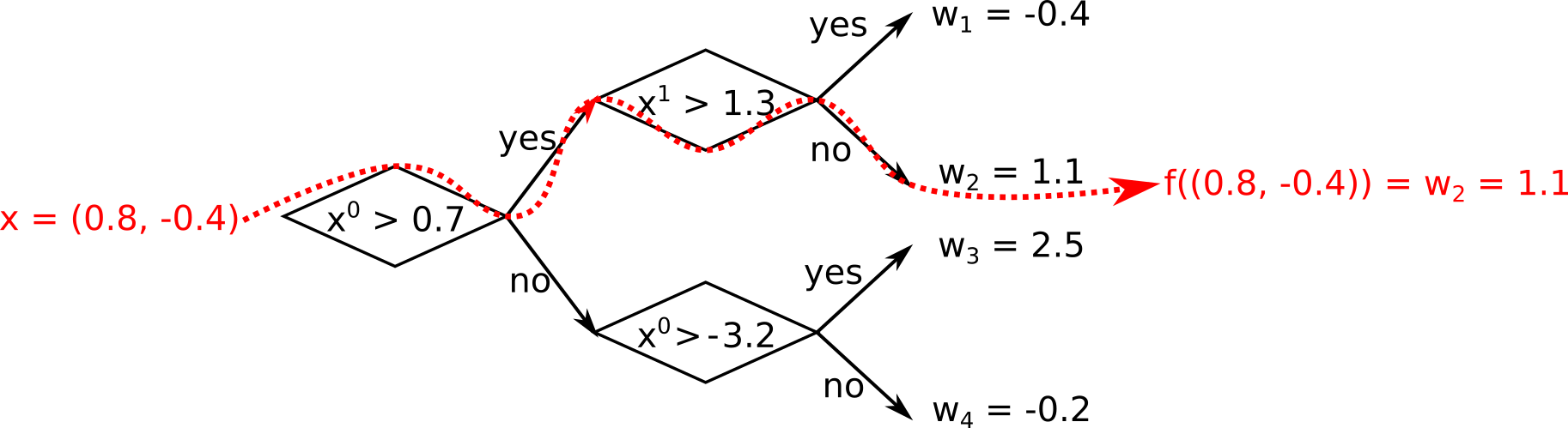}
  \caption{Example of regression tree with four leaves. In red, application of the function $f(x)$ associated with the tree to an input $x$.} \label{fig:regrtree}
\end{figure}

Regression trees are a type of decision tree. A decision tree is a diagram that
recursively partitions data into subsets, based on the binary (true/false)
conditions located at the nodes of the tree. The final subsets in which the
data are classified are called ``leaves''. A decision tree with $T$ leaves is
formally a function $q:\mathbb{R}^n\to \{1,2,\ldots,T\}$ which associates to
each data point $x \in \mathbb{R}^n$ its leaf $q(x)$. A regression tree
assigns to each leaf $i$ a real number $w_i \in \mathbb{R}$. The regression
tree therefore defines a function $f:\mathbb{R}^n\to\mathbb{R}$, given by
\begin{eqnarray}
  f(x) = w_{q(x)}\,.
\end{eqnarray}
An example of a regression tree with four leaves is depicted in
Fig.~\ref{fig:regrtree}. In
practice, a single tree is not general enough to reproduce an arbitrary
function. For this reason, we consider instead an ensemble of $K$ regression
trees $\mathcal{F} = \{f^{(1)},f^{(2)},\ldots,f^{(K)}\}$. The ensemble defines
a function $\phi:\mathbb{R}^n \to \mathbb{R}$,
\begin{equation}
  \phi(x) = \sum_{i=1}^K f^{(i)}(x) = \sum_{i=1}^K w^{(i)}_{q(x)}\,.
\end{equation}

The function $\phi(x)$ will represent the approximation for the log-likelihood
function. It will be calculated using supervised learning, that is, the
trees are obtained from a dataset $\mathcal{D}=\{(x_i, y_i)\}$ where $x_1,
  \ldots x_N \in \mathbb{R}^n$ are the inputs and $y_1,\ldots y_N \in \mathbb{R}$
are the pre-computed outputs for each input. In our case, the input data will be
of the form $x_i = (C_i, \alpha^\ell_i, \beta^\ell_i, \alpha^q_i, \beta^q_i)$,
and the outputs will be $y_i = \log L(x_i)$.

In order to train the model from the dataset, we need to define an objective
function $\mathcal{L}[\phi]$ that measures how well the model fits the data,
\begin{equation}
  \mathcal{L}[\phi] = \sum_i l(\phi(x_i), y_i) + \sum_k \Omega(f^{(k)})\,,
\end{equation}
which has two components:
\begin{itemize}
\item The loss function $l(\phi(x_i), y_i)$ is a differentiable function that
measures the similarity between the true output $y_i$ and its approximation $\phi(x_i)$. We use as loss function the mean absolute error, $l(\phi(x_i), y_i) = |\phi(x_i) - y_i|$.
\item The function $\Omega$ is the regularization term, that penalizes the
complexity of trees, that is, trees with many leaves or with large $||w||$.
The purpose of the regularization is to prevent over-fitting, that is,
the model learning ``by heart'' the training data and being unable to extrapolate from them.
\end{itemize}

The ensemble is constructed in an iterative way, starting from one single
tree $f^{(0)}$ that contains just one leaf. At the step $t$ of the iteration,
the tree $f^{(t)}$ is obtained by splitting one of the leaves of the tree
$f^{(t-1)}$ into two leaves; the splitting is determined by the optimization
of the objective function. In order to prevent over-fitting, the shrinkage
technique is used, that scales newly added weights by a factor $\eta<1$, similar
to the learning rate in other Machine Learning algorithms.

Once we have an approximation of the log-likelihood function, we put it to
use and generate new samples of datapoints $x_i = (C_i, \alpha^\ell_i, \beta^\ell_i, \alpha^q_i,
  \beta^q_i)$. We use a Montecarlo algorithm to produce the data distributed
according to the $\chi^2$ distribution of the fit. At each step of the
Montecarlo algorithm, a new tentative $x_i$ is proposed, which is accepted
if the ratio of its probability divided by the probability of the best fit
point is greater than a random number $u$ distributed uniformly in the interval
$[0,1]$, and rejected otherwise. Expressed in terms of the logarithms of the
likelihood function instead,
\begin{equation}
\log L(x_i) > \log L_\mathrm{bf} + \log u\,,
\end{equation}
where $L_\mathrm{bf}$ is the likelihood of the best fit.
This algorithm requires many calls to the likelihood function, which are
computationally very tasking, and most of the proposed points are rejected.
As a way to ease the burden, we use the approximated log-likelihood $\phi(x_i)$
instead of the true function.

We can asses the importance of each parameter in the Machine Learning
approximation at any point of the generated samples by using SHAP
(SHAPley Additive exPlanation) values~\cite{Lundberg:2017,2018arXiv180203888L}.
SHAP values are based in Lloyd Shapley's work on game
theory~\cite{Shapley+2016+307+318}, who won the Nobel Prize in Economics for it
in 2012. 

\begin{figure}
\centering
\includegraphics[width=0.75\textwidth]{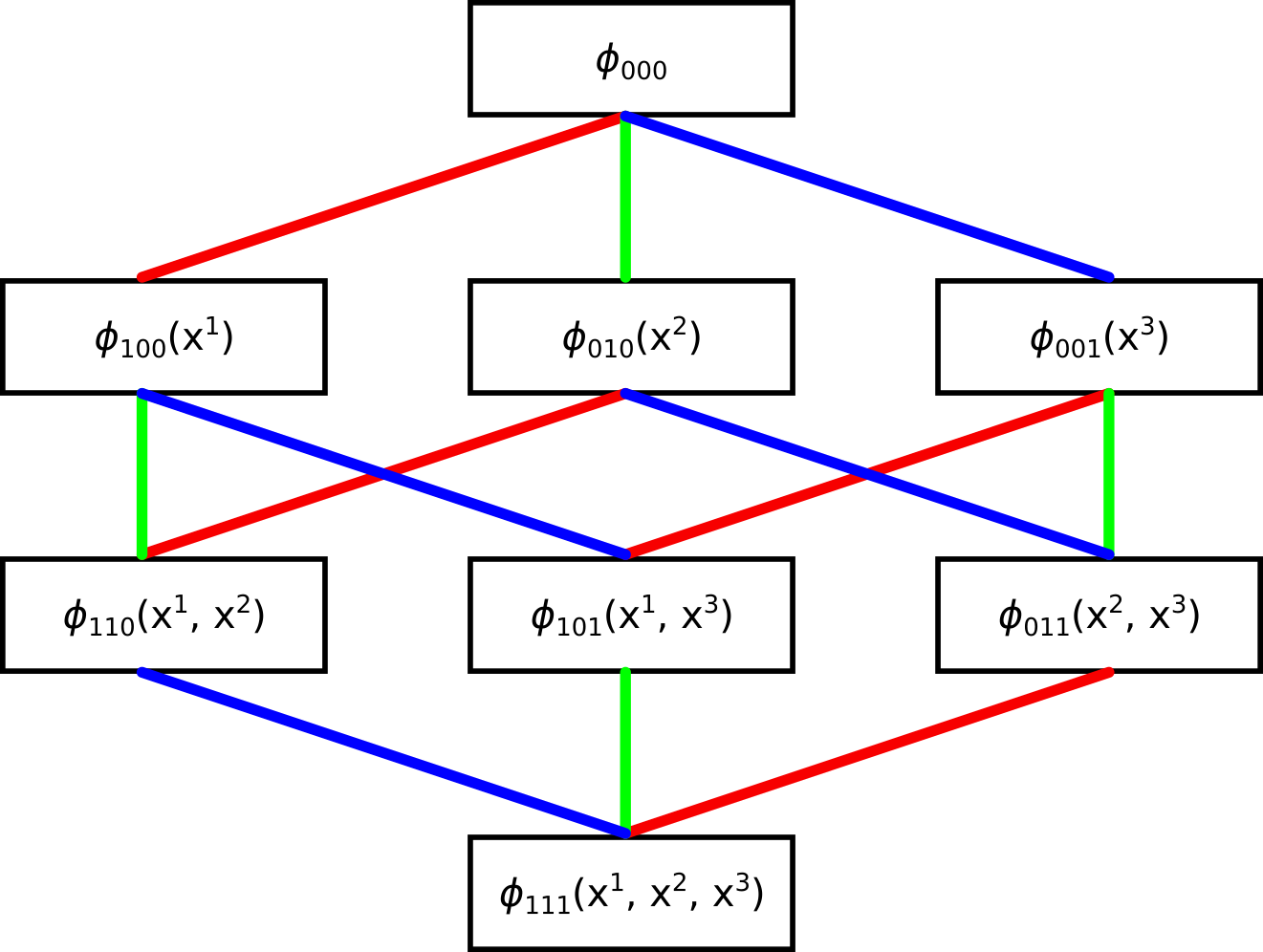}
\caption{Prediction models that would we necessary to train for three features in order to calculate the SHAP values. The edges represent the marginal contributions for each feature: in red for $x^1$, green for $x^2$ and blue for $x^3$.}\label{fig:shapgraph}
\end{figure}

The SHAP values are designed with three properties in mind:
\begin{itemize}
  \item \textbf{Local accuracy:} The sum of the SHAP values is equal to the model prediction.
  \item \textbf{Missingness:} If any feature is missing, its SHAP value is zero.
  \item \textbf{Consistency:} If the model is changed so any feature has larger impact, its SHAP value will increase.
\end{itemize}

Given a model $\phi(x)$, the SHAP trains $2^n$ new models $\phi_z(x)$
for $z\in\{0,1\}^n$ binary vectors. The model $\phi_z(x)$ contains the
feature $x^{(i)}$ only if $z^{(\alpha)}=1$, while that feature is
ignored when training if $z^{(\alpha)}=0$. The marginal contribution
$\phi_{z'}(x_i) - \phi_z(x_i)$ for two models differing only in the
presence of one feature (i.e. $z^{(\alpha)} = 0$, ${z'}^{(\alpha)} = 1$
and $z^{(\beta)} = {z'}^{(\beta)}\ \forall\ \beta\neq \alpha$), gives
the importance of adding the feature $\alpha$ to the model $z$. The SHAP
value for the feature $\alpha$ in the point $x_i$ is just the weighted
average of all marginal contributions, with the weight given by a
combinatorial factor. An example is depicted in
Fig.~\ref{fig:shapgraph}. The prediction without any features
$\phi_{0\cdots0}$ is simply the average of the values $y_i$ in the
dataset, and acts as a base value common for all $x_i$. 

Finally, we will analyze the correlations between the points in the
generated samples, in order to understand the physical relations caused
by NP.

\subsection{Procedure and results}

In the first place we create a sample of 5000 parameter points and their
likelihood using the traditional algorithm. We discard the points with
$\Delta\chi^2_\mathrm{SM} < 20$, retaining 3760 points. 

\begin{figure}
  \centering
  \begin{tabular}{cc}
    \includegraphics[width=0.45\textwidth]{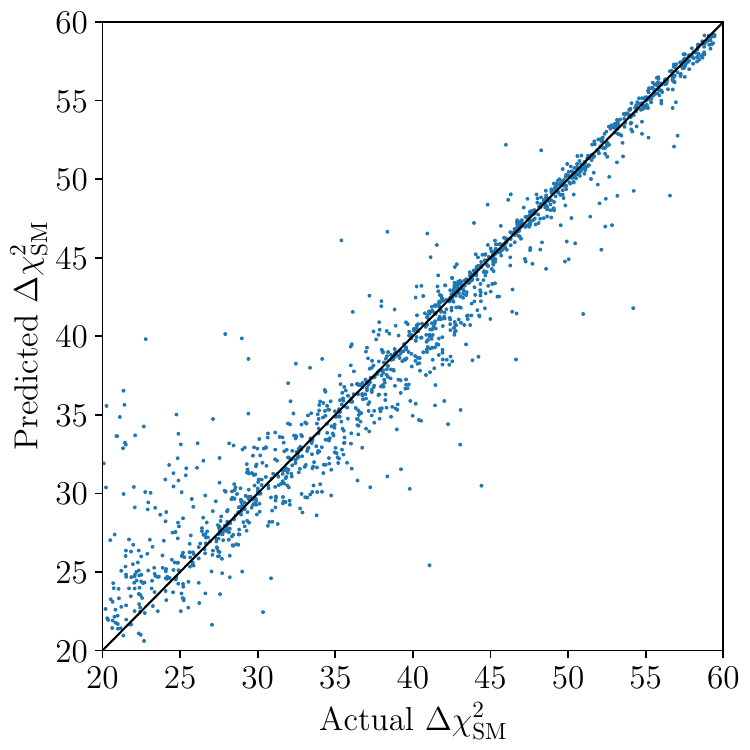} &
    \includegraphics[width=0.45\textwidth]{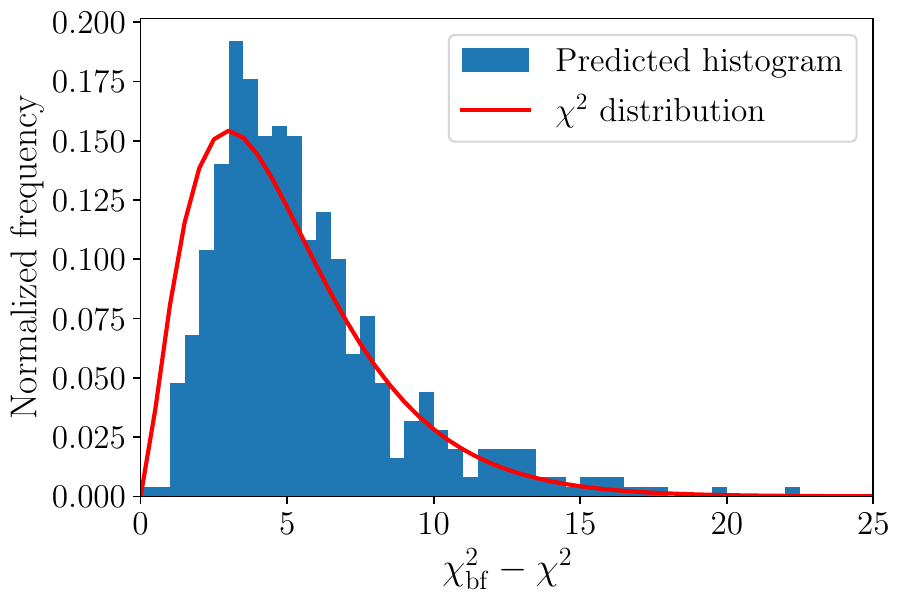}
    \\
    (a)  & (b) \\
  \end{tabular}
  \caption{Predictor performance: (a) Regression of the predicted values
    of $\Delta\chi^2$ compared to the real ones in the validation
    dataset. (b) Histogram of the Montecarlo points generated using the
    Machine Learning algorithm: blue bins for the predicted distribution
    and orange bins for the actual distribution. } 
  \label{im:predictor}
\end{figure}
We train a Machine Learning predictor using the pre-computed sample.
We used the XGBoost (eXtreme Gradient Boosting)
algorithm~\cite{Chen:2016}, implemented by the Python
package \texttt{xgboost}. We split the sample in two parts, 75\% of
the points for the training and 25\% points for the validation of the
model. The algorithm uses a learning rate of 0.05 and 1000 estimators,
allowing early stopping at 5 rounds.
The performance of the Machine Learning predictor can be seen in
Figure~\ref{im:predictor} (a). 
The horizontal axis represents the actual value of the $\Delta\chi^2_\mathrm{SM}$ for each point of the validation dataset, computed using the full \texttt{flavio} and \texttt{smelli} code, with the best fit point found in Section~\ref{sec:Fits} corresponding to the maximum value. The vertical axis represents the predicted value for the same points obtained using the Machine Learning Montecarlo algorithm.
The predicted values for the $\Delta
\chi^2_\mathrm{SM}$ reproduce their actual values, with a Pearson
regression coefficient $r = 0.970$ and Mean Absolute Error of $0.719$ in
the validation dataset. The agreement between the predicted and actual
values is specially good for parameters near the best fit point ($\Delta
\chi^2_\mathrm{SM} > 55 $). 

Finally we implement the Montecarlo algorithm: we generate random
points $\vec{C}$ near the best fit. The predictor produces an
approximation of the $\Delta\chi^2$, and therefore, also an
approximation of the logarithm of the likelihood,
$\log\tilde{L}(\vec{C})$. The point is accepted if this approximation
verifies the Montecarlo condition
\begin{equation}
  \log\tilde{L}(\vec{C}) > \log L_\mathrm{bf} + \log u\,,
\end{equation}
where $L_\mathrm{bf}$ is the likelihood of the best fit and $u$ is a
number randomly chosen from an uniform distribution in the interval
$[0,1)$. To check if the Machine Learning Montecarlo algorithm can actually
reproduce the $\chi^2$ distribution, we generate a sample of 1000
points. The histogram for the predicted  values of the
$\chi^2$ is plotted in Figure~\ref{im:predictor} (b). The histogram
follows the general shape of the $\chi^2$ distribution, although there is
an excess of points near the best fit and a deficit of points in the
region of low likelihood. 

In order to understand how each parameter affects the prediction of the
likelihood, we use the SHAP
values as described above. Remember that once we have an approximation
of the log-likelihood function, we put it to use
generating new samples of datapoints $x_i = (C_i, \alpha^\ell_i, \beta^\ell_i, \alpha^q_i,
  \beta^q_i)$, and we use a Montecarlo algorithm to produce the data
distributed according to the $\chi^2$ distribution of the fit.
Table~\ref{tab:SHAP_bf} contains an example of the SHAP values
for $\log L$ at the best fit point. According to the Machine Learning
model, the values of $C$ and $\alpha^\ell$ and $\beta^q$ have the larger 
impact in the Machine Learning prediction. 
\begin{table}
\centering
\begin{tabular}{|*{8}{c|}}\hline
Base & \multicolumn{5}{c|}{SHAP value for} & Final & Actual \\ \cline{2-6}
value & $C$ & $\alpha^\ell$ & $\beta^\ell$ & $\alpha^q$ & $\beta^q$ & prediction & $\log L$ \\\hline
40.552 & 3.599 & 3.355 & 3.384 & 2.375 & 3.914 & 57.180 & 58.844 \\\hline
\end{tabular}
\caption{SHAP values and Machine Learning prediction for the best fit point.}\label{tab:SHAP_bf}
\end{table}
\begin{figure}
\centering
\includegraphics[height=55mm, trim= 0in 0in 1.3in 0in, clip]{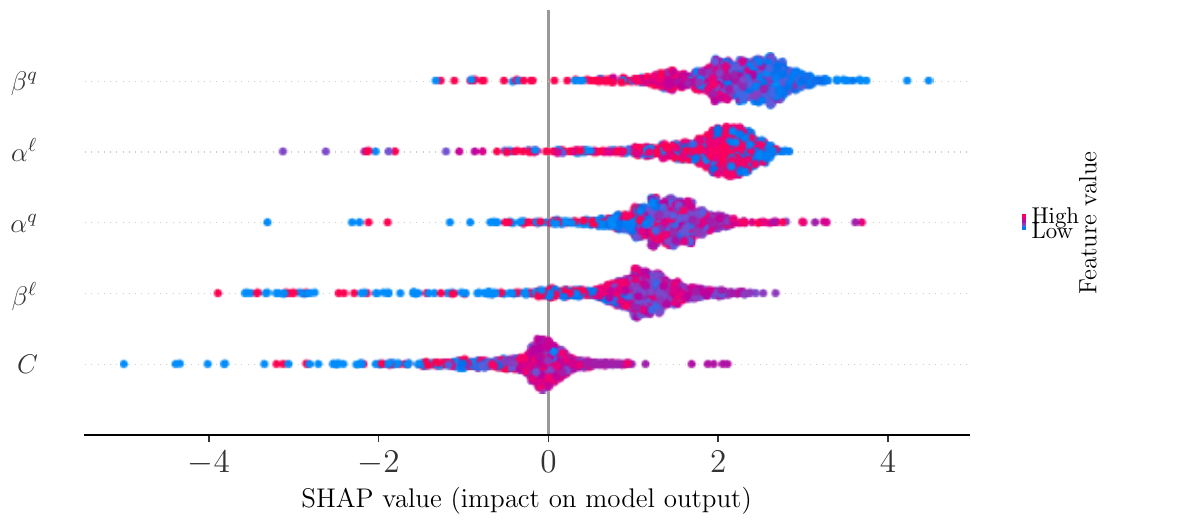}
\includegraphics[height=55mm]{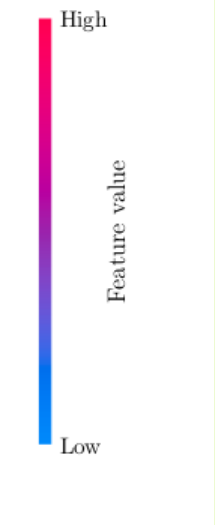}
\caption{Distribution of the SHAP values for each parameter in a sample
  of 10000 generated points.}
\label{im:mean_SHAP}
\end{figure}
\begin{figure}
\centering
\begin{tabular}{ccc}
  \includegraphics[width=0.3\textwidth]{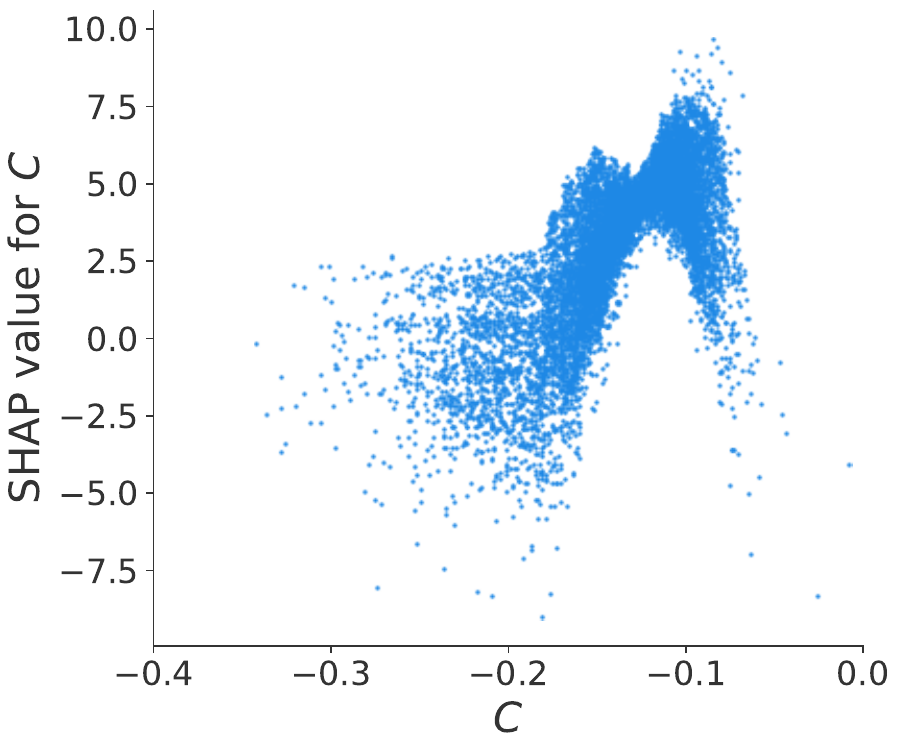}
  & \includegraphics[width=0.3\textwidth]{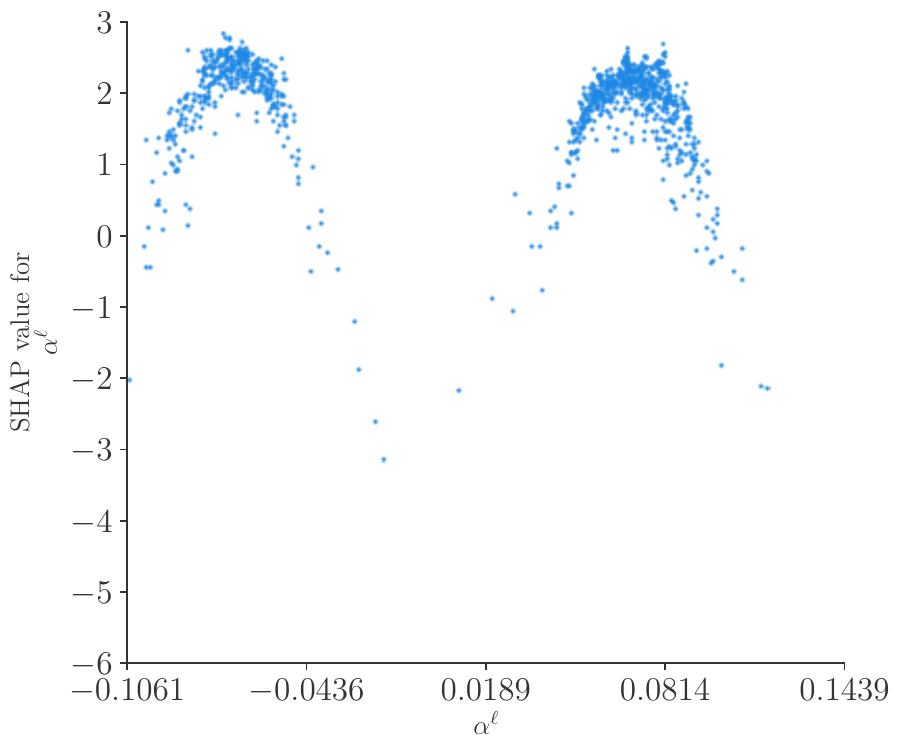}
  & \includegraphics[width=0.3\textwidth]{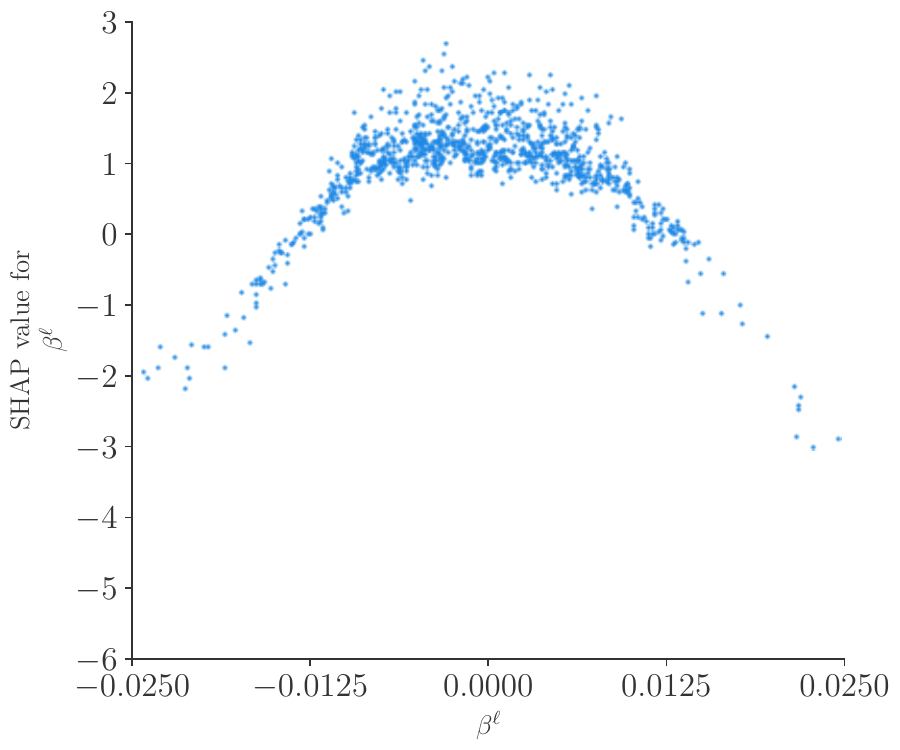} \\
(a) & (b) & (c) \\
& & \\
\end{tabular}
\begin{tabular}{cc}
\includegraphics[width=0.3\textwidth]{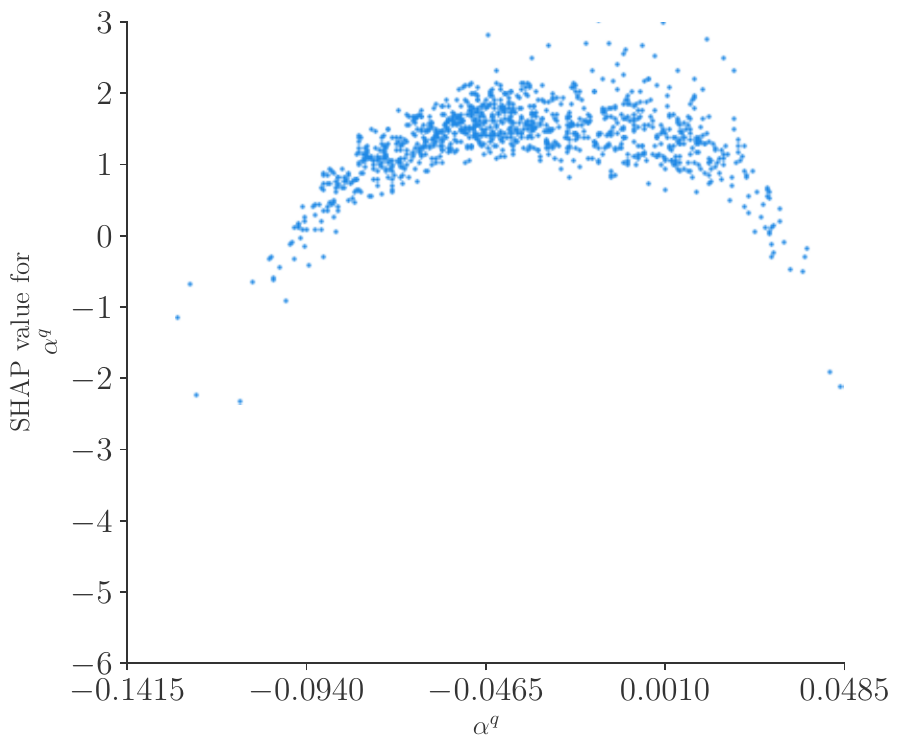} & \includegraphics[width=0.3\textwidth]{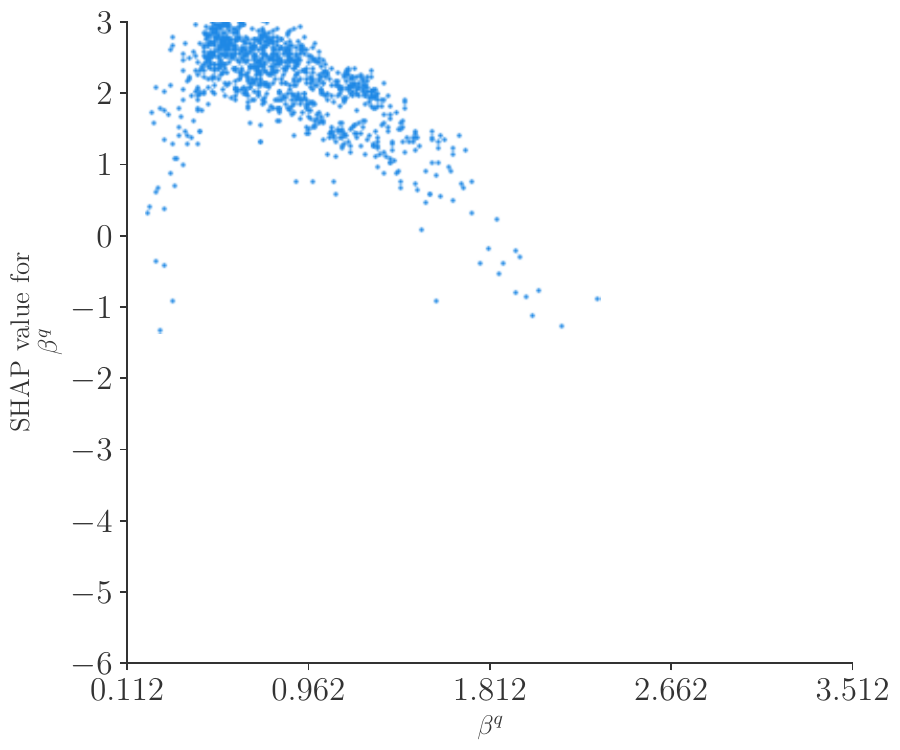} \\
(d) & (e)
\end{tabular}
\caption{SHAP values for the parameters of the fit at the sample of
  10000 generated points.}
\label{im:SHAP_parameters}
\end{figure}

Figure~\ref{im:mean_SHAP} shows the impact of each parameter to the final
prediction, measured as the mean of the absolute values of their SHAP
values across a sample of 10000 Montecarlo points. The SHAP 
values allow us to quantify the relative importance of each parameter
in the fit. The parameters $\beta^q$ and $\alpha^\ell$ have the largest
contribution and $\beta^\ell$ and $C$ contribute the less.
This result is in disagreement with the assumption of
Scenario I regarding the parameter $\alpha^\ell$ describing the mixing to the
first generation of leptons.
Therefore, the obtained result is in agreement with the
previous section, where we already concluded that the mixing with the
first generation were necessary in order to describe both anomalies
simultaneously. On the other hand we see that, while moderate values of $\beta^\ell$ are 
relatively unimportant compared to the other mixing parameters, extreme values 
have a large negative impact in the prediction for the likelihood.

We calculate the SHAP values for the logarithm of the likelihood at each
point of the Montecarlo sample. In this way, we can determine how each
parameter contributes to the fit, as shown in
Figure~\ref{im:SHAP_parameters}. We can compare these SHAP values with
Figure~\ref{im:likelihood_alphabeta}, where only one parameter was changed
at a time. Then, it is clear the agreement between the results obtained by the
Machine Learning Montecarlo algorithm proposed in this work and the ones
obtained by following the RG equations. Therefore,
we can conclude that the SHAP values reproduce correctly the
general features of the fit.

The above results show that the Machine Learning Montecarlo algorithm
can be very useful in this kind of analysis, being able to reproduce
the results obtained in the previous section in a shorter time.
We can conclude that the machine
learning, made jointly with the SHAP values, constitute a suitable
strategy to use in complex fitting problems with large dimensionalities
and complicated constraints, where a direct evaluation is too
time-consuming.

\subsection{Correlations between observables}

In order to check our Machine Learning procedure, we now discuss
on the agreement of the results obtained by the Machine Learning Montecarlo algorithm
that we have proposed and the ones obtained by using the RG equations defined
as given in Section~\ref{sec:EFT}. 

\begin{figure}
\centering
\includegraphics[width=0.6\textwidth]{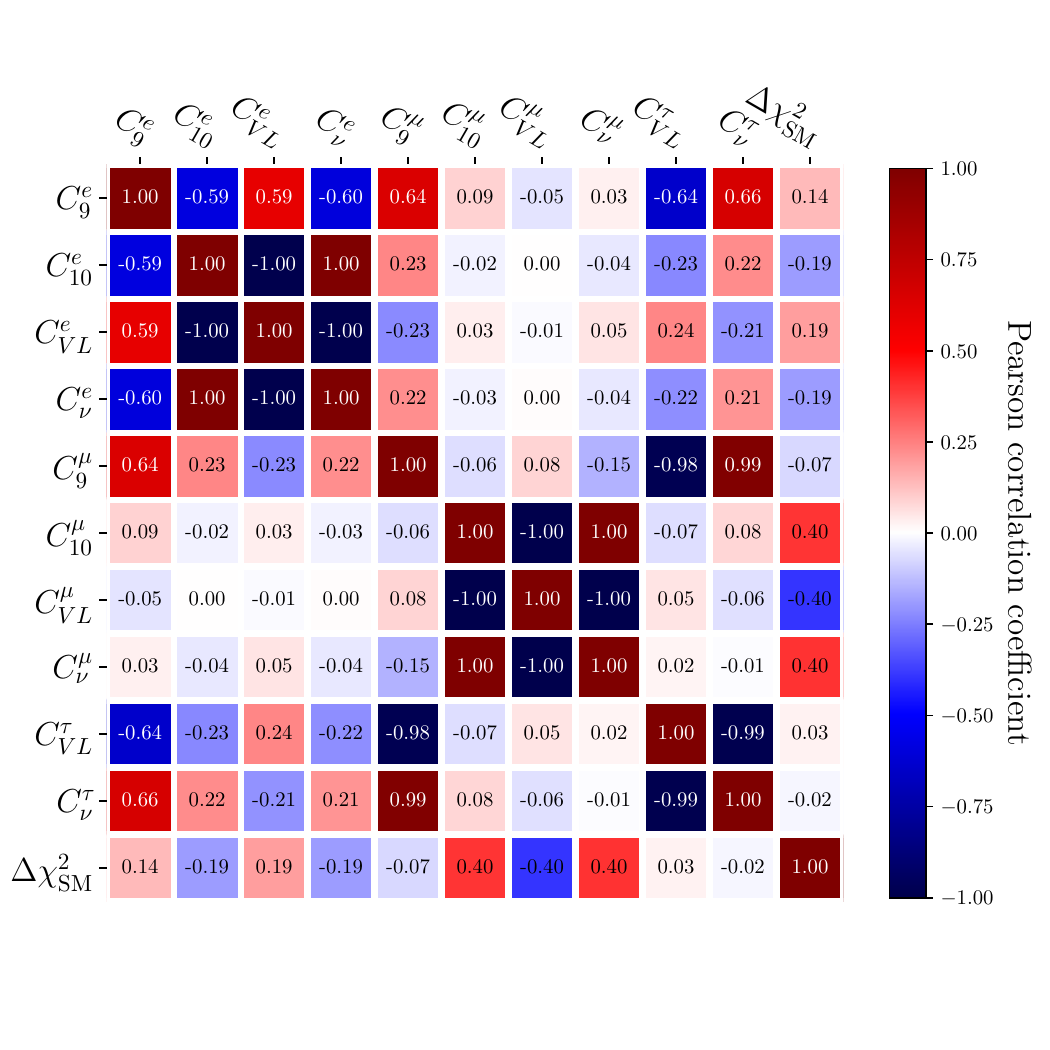}
\caption{Matrix of Pearson correlation coefficients between semileptonic
  WET Wilson Coefficients in the sample of 15000 points in parameter
  space.}
\label{im:coeffcorr} 
\end{figure}

The Lagrangian in Eq.~\eqref{eq:LagLambda} exhibit a flavour structure,
given by the $\lambda$ matrices, relating the different entries of the
tensor of Wilson coefficients $\Clq^{ijkl}$. Under the RG evolution and
matching, this flavour structure is imprinted in the
WET Lagrangian in Eq.~\eqref{eq:lagWET}, and therefore in the
related observables. 
Using the Machine-Learning Montecarlo algorithm described in the
previous section, we generate a sample of 15000 points in parameter
space around the best fit point. In each point we run the RG equations
down to the electroweak scale, perform the matching with the WET, and
run the RG equations again down to $\mu=m_b$. We compute the
correlations between the semileptonic $b\to s$ and $b\to c$ coefficients
$C_9$, $C_{10}$, $C_{VL}$ and $C_\nu$ for the different lepton
generations. Figure~\ref{im:coeffcorr} shows the matrix of Pearson
coefficients describing linear correlations between the WET Wilson
Coefficients. In the electron sector, $C_{10}^e$, $C_{VL}^e$
and $C_\nu^e$ show strong correlations close to $\pm 1$. In the muon
sector, $C_{10}^\mu$, $C_{VL}^\mu$ and $C_\nu^\mu$ are also correlated
between them, however they are linearly independent of
$C_9^\mu$. Instead, $C_9^\mu$ is correlated with the tau coefficients
$C_{VL}^\tau$ and $C_\nu^\tau$, and to a lesser extent to $C_9^e$. 
\begin{figure}
\centering
  \includegraphics[width=0.6\textwidth]{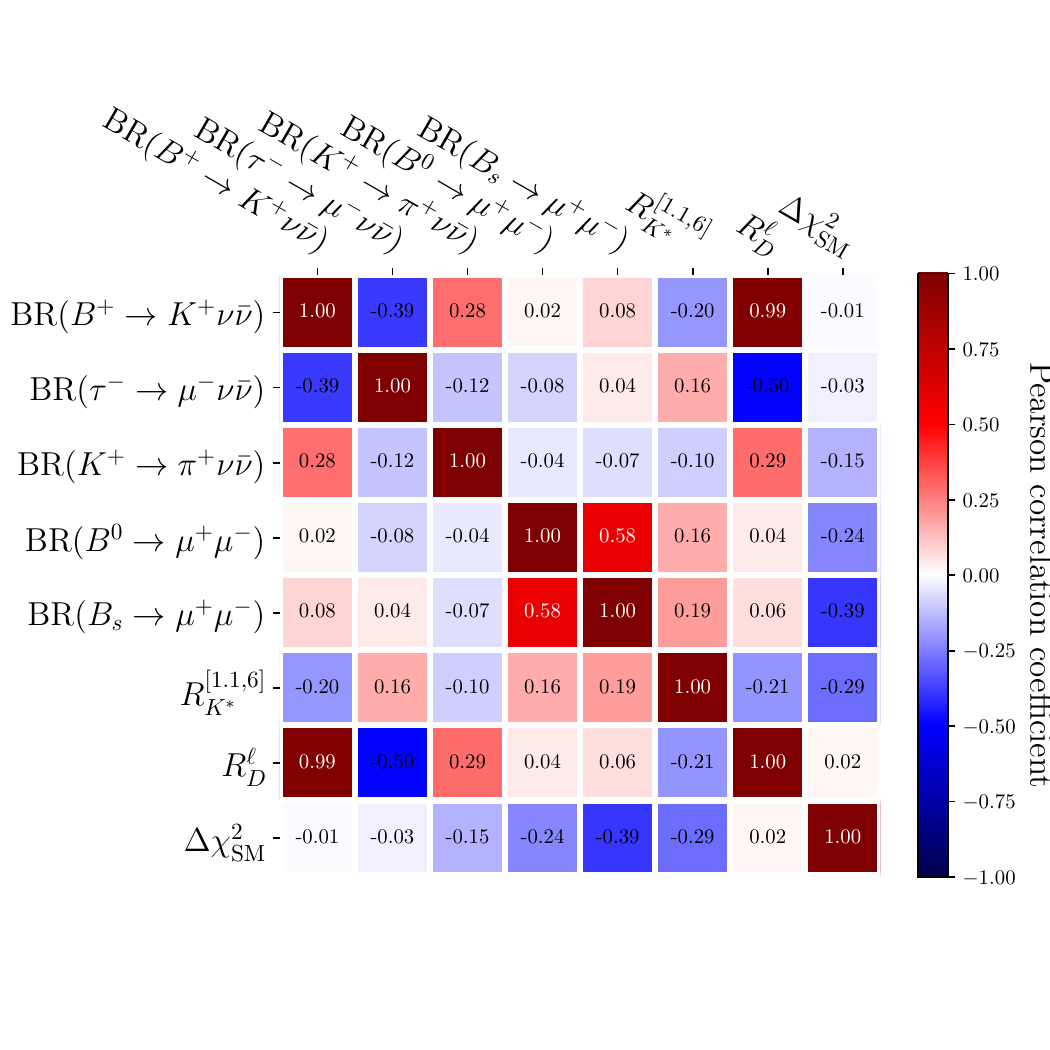} 
\caption{Matrix of Pearson correlation coefficients for selected observables in the 15000 points sample.}\label{im:obscorr}
\end{figure}

The correlations that we have found are consistent with the results of
RG evolution and matching in Eq.~\eqref{eq:matching}. In the case of the
electron sector, the $C_{10}^e$, $C_{VL}^e$
and $C_\nu^e$ coefficients are all proportional to the product
$C \lambda^q_{23} \lambda^\ell_{11}$ appearing in the tree-level contribution.
Analogously in the muon sector 
$C_{10}^\mu$, $C_{VL}^\mu$ and $C_\nu^\mu$, depend on $C\lambda^q_{23}\lambda^\ell_{22}$
and in the tau sector $C_{VL}^\tau$ and $C_\nu^\tau$ depend on $C\lambda^q_{23}\lambda^\ell_{33}$. 
The coefficient $C_9^\mu$ is not correlated to the rest of the muonic coefficients
because it is dominated by the loop-level contribution $C_9^\mathrm{loop}$,
which depends on the product $C\lambda^q_{23}$. 
The coefficient $C_9^e$ receives sizeable contributions both from the tree-level
and the one-loop terms, and consequently shows a mild correlation with $C_9^\mu$
and a total correlation with the combination $C_9^\mu - C_{10}^e$. 
Lastly, there is a perfect correlation of $\pm1$ between $C_9^\mu$ and the tau
coefficients, which is caused by the fact that $\lambda^l_{33} = 0.994\pm 0.001$
is almost constant, so $C\lambda^q_{23} \lambda^l_{33} \approx C\lambda^q_{23}$.
We can therefore conclude that the obtained data is in agreement with the
arrangement of Wilson coefficients presented in Eq.~\eqref{eq:matching}. 

Besides, in the same sample of 15000 points, we determine the predictions of our
model for several selected observables of various flavour sectors, with
large pull differences between the SM and Scenario II predictions:
$R_{K^*}^{[1.1,6]}$ (observable 12 in the table presented in
Appendix~\ref{app:pulls_scbII}), $\mathrm{BR}(B^+\to K^+\nu\bar{\nu})$  (observable 94)
and $\mathrm{BR}(B_s\to\mu^+ \mu^-)$ (observable 45) from $b\to s$ decays,
$R_D^\ell$ (observable 77) from $b\to c$ decays,
$\mathrm{BR}(B^0\to \mu^+\mu^-)$ (observable 245) from $b\to d$ decays,
$\mathrm{BR}(K^+\to \pi^+ \nu \bar{\nu})$ (observable 406) from $s\to d$ decays
that has a great impact in the fit value of $\alpha^q$, and the tau decay
$\mathrm{BR}(\tau^-\to \mu^-\nu\bar{\nu})$ (observable 37).
The correlation matrices are depicted in Figure~\ref{im:obscorr}.

\begin{figure}
\centering
\includegraphics[width=0.7\textwidth]{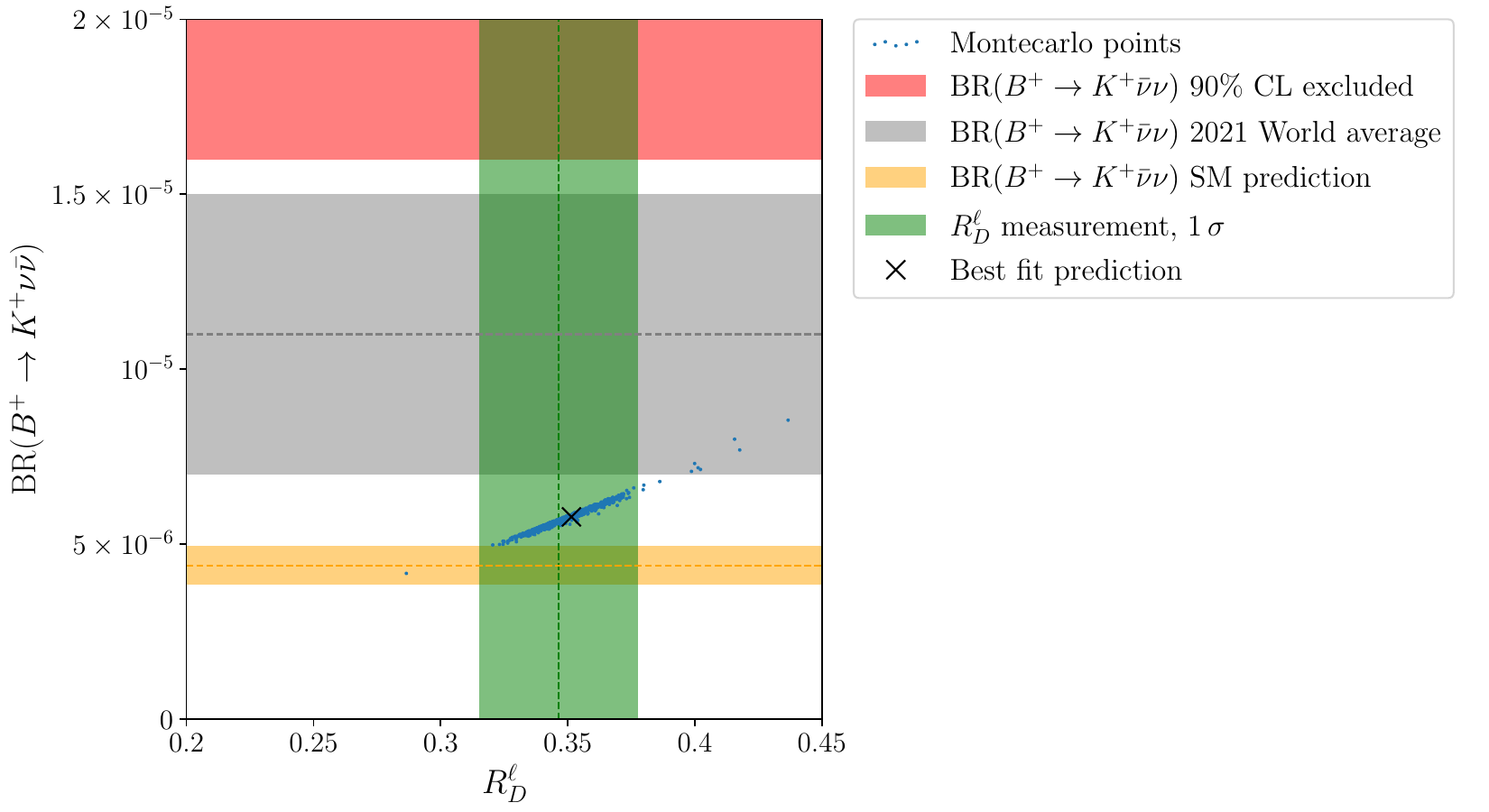}
\caption{Predictions for the observables $R_{D^*}^\ell$ and
  $\mathrm{BR}(B^+\to K^+ \nu \bar{\nu})$ in the generated sample. The
  green vertical band corresponds to the $R_{D^*}^\ell$
  \cite{Amhis:2019ckw} measurement, the red horizontal band to the 90\%
  CL excluded region for $\mathrm{BR}(B\to K^* \nu \bar{\nu})$
  \cite{Grygier:2017tzo}, the grey band to the 2021 world average
  obtained by Belle II \cite{Dattola:2021cmw} and the yellow horizontal
  band to its SM prediction.}\label{im:RD_BKnunu} 
\end{figure}

From the above results, it is clear that
the observables $R_{D^*}^\ell$ and $\mathrm{BR}(B^+\to K^+ \nu \bar{\nu})$
show an almost-perfect correlation. Then, predictions for these two
observables in the generated sample are shown in
Figure~\ref{im:RD_BKnunu}. The green vertical band in this figure
corresponds to the $R_{D^*}^\ell$ measurement~\cite{Amhis:2019ckw},
the red horizontal band to the $90\%$ CL excluded region for
$\mathrm{BR}(B\to K^* \nu \bar{\nu})$~\cite{Grygier:2017tzo} and
the grey band to the 2021 world average obtained by Belle II~\cite{Dattola:2021cmw}.
The yellow horizontal band summarizes the SM prediction.
The obtained values of Montecarlo points and the best fit prediction
of our computations are also included.
It is important to stress that $R_{D^*}^\ell$ depends on the Wilson
coefficient $C_{VL}^\tau$, and $\mathrm{BR}(B^+\to K^+ \nu \bar{\nu})$ on
$C_\nu^\tau$, and both of them are proportional to the product $C
\lambda^q_{23}\lambda^\ell_{33}$. This is in contrast with the conclusions
of~\cite{Browder:2021hbl}, where several leptoquark scenarios coupling
to right-handed neutrinos did not find a significant correlation between
both observables. Even if the correlation is strong, the
prediction for the $B^+\to K^+\nu\nu$ decay remains compatible with the
90\% confidence level (CL), $\mathrm{BR}(B\to K^+ \nu \nu)<1.6\times
10^{-5}$~\cite{Grygier:2017tzo}, for the whole range of experimentally-compatible values
of $R_{D^*}^\ell$. The world average for the branching ratio obtained by
Belle II~\cite{Dattola:2021cmw} (not included in our numerical analysis)
shows an enhancement of a factor of $2.4\pm 0.9$ compared to the SM
prediction~\cite{Browder:2021hbl}. While our data is in tension with this
world average, it is an encouraging sign of a possible interplay between
$\RDp$ and $\mathrm{BR}(B^+\to K^+ \nu \bar{\nu})$. Future experimental
results from Belle II will further clarify the  situation.

It is worth stressing that the observable $R_{K^*}$ displays a moderate correlation with
$R_{D^*}^\ell$ and $\mathrm{BR}(B^+\to K^+ \nu \bar{\nu})$, caused by the
relation of the Wilson coefficient $C_9^\mu$ with $C_{VL}^\tau$ and
$C_\nu^\tau$. On the other hand, $R_{K^*}$ shows almost no correlation
to $\mathrm{BR}(B_s\to \mu^+ \mu^-)$, even though both observables
depend on $C_{10}^\mu$. This is a result that sets us apart from many NP
models that impose the relation $C_9^\mu = -C_{10}^\mu$. 

Finally, none of the selected observables display a large correlation to the
goodness of fit measured by $\Delta \chi^2$. This is a sign that there is not a
single observable dominating the fit, and reaffirms that global fits are in
fact a necessity on the analysis of flavour anomalies.

\section{Connection to leptoquark models}\label{sec:LQ}

In this section we discuss the phenomenological implications of
our assumptions in the vector leptoquark model.
The vector leptoquark $U_1 = (\mathbf{\bar{3}},\mathbf{1})_{2/3}$
couples to left-handed and right-handed fermions as 
\begin{equation}
\mathcal{L}  = x^{ij}_L \bar{q}_i \gamma_\mu U_1^\mu \ell_j + x_R^{ij} \bar{d}_{Ri} \gamma_\mu U_1^\mu \ell_{Rj} + \mathrm{h.c.}
\end{equation}
An $U_1$ leptoquark with mass $M_U$, when matched with the SMEFT at the
scale $\Lambda$, contributes to the following Wilson coefficients~\cite{delAguila:2010mx}: 
\begin{align}
C_{\ell q(1)}^{ijkl} = C_{\ell q(3)}^{ijkl} &= \frac{- \Lambda^2}{2 M_U^2} x_L^{li} x_L^{kj*}\,, \nonumber \\
C_{ed}^{ijkl} = -\frac{1}{2} C_{qde}^{ijkl} &= \frac{-\Lambda^2}{M_U^2} x_R^{li} x_R^{kj*}\,.
\end{align}

Our model does not include couplings to right-handed leptons in the
interaction Lagrangian, and therefore all the $x_R$ couplings are set to
zero. The left-handed couplings $x_L$ are related to the parameters of
the Lagrangian \eqref{eq:LagLambda} according to  
\begin{align}
|x_L^{ji}|^2 = -\frac{2M_U^2}{\Lambda^2} C \lambda_{ii}^\ell \lambda_{jj}^q \,, \nonumber \\
\mathrm{Arg}(x_L^{ji}) = \mathrm{Arg}(\lambda^q_{j3}) - \mathrm{Arg}(\lambda^\ell_{i3}) + \theta\,,
\label{eq:U1matching}
\end{align}
where $\theta$ is a free global complex phase. Since the rotation
matrices $\lambda$ are hermitian ($\lambda^\ell_{ii}$ and
$\lambda^q_{jj}$ are real and positive), we need $C_1 = C_3$ to be a
real negative number. This condition is fulfilled in both Scenarios I
and II. 

Without loss of generality we set $\theta = 0$.
The mass of the leptoquark is chosen to be $M_U = 1.5\ \mathrm{TeV}$,
the lowest mass not excluded by direct searches~\cite{CMS:2018bhq}.
The results in Scenario I correspond to 
\begin{equation}
x_L = \begin{pmatrix}
0 & 0 & 0\\
0 & 1.6\times 10^{-9} & 0.452\\
0 &  2\times 10^{-9}  & 0.580
\end{pmatrix}\,,
\end{equation}
and the results of Scenario II are
\begin{equation}
x_L = \begin{pmatrix}
-3.18\times10^{-3} & 3.00\times 10^{-10}  & -0.0431 \\
0.0318 & -3.00\times 10^{-9} &  0.430\\
0.0446  & -4.20\times 10^{-9}  & 0.604
\end{pmatrix}\,.
\end{equation}

In both scenarios, the most important coupling are $x_L^{23}$ to second generation quarks and third generation leptons, and $x_L^{33}$ to third generation quarks and leptons. A similar leptoquark model has been proposed previously, as scenario RD2A in \cite{Bhaskar:2021pml} as a solution for the $\RDp$ anomaly. The advantage of our proposal is that the inclusion of small non-zero values of the couplings $x_L^{21}$ and $x_L^{31}$ is able to explain the $\RKp$ anomalies at the same time. The values of $x_L^{23}$ and $x_L^{33}$ are compatible with the exclusion limits set in \cite{Bhaskar:2021pml}.

Other leptoquark models do not retain the $\Clqo = \Clqt$ condition
\cite{delAguila:2010mx,Bhattacharya:2016mcc}, and therefore produce
large contributions to the $B \to K^{(*)}\nu \bar{\nu}$ decays. That is
the case of the scalar $S_3 = (\mathbf{\bar{3}}, \mathbf{3})_{1/3}$,
that predicts $\Clqo = 3 \Clqt$, and the vector $U_3 =
(\mathbf{\bar{3}}, \mathbf{3})_{2/3}$, where $\Clqo = - 3 \Clqt$. The
scalar $S_1 = (\mathbf{\bar{3}}, \mathbf{1})_{1/3}$ is even less suited,
as it predicts $\Clqo = - \Clqt$, which would result in no NP
contributing to $b\to s \ell^+ \ell^-$ at all. New vector bosons $W'$
and $Z'$ would also be in conflict with the $B \to K^{(*)}\nu \bar{\nu}$
decays, as they predict $\Clqo = 0$ while \Clqt has a non-zero value. 

\subsection{A simplified model}
\label{sec:simplifiedmodel}

In this section, we will propose a simplified $U_1$ leptoquark model, depending 
only on two coupling constants, that reproduces the numerical results that we 
have obtained in section~\ref{sec:Fits}.
This scenario implies that the NP contributions to
the $C_9^{e}$ and $C_9^{\mu}$ Wilson
coefficients are of the same order, but the ones to $C_{10}^{e}$ is two
orders of magnitude larger than
to $C_{10}^{\mu}$.

In the quark sector, we assume that the leptoquark does not interact with the 
first generation quarks, and that it interacts equally with second and third 
generation quarks. The rotation matrix corresponding to this assumption has 
elements $(\hat{U}_q)_{31} = 0$ and $(\hat{U}_q)_{32} = (\hat{U}_q)_{33} = \frac{1}{\sqrt{2}}$. 
The parameters of the mixing matrix, $\alpha^q=0$ and $\beta^q=1$, are 
compatible with the results of the fit obtained in Table~\ref{tab:rotation}.

For the leptonic sector, we assume that the leptoquark interacts differently 
with each generation, being the interaction with the second generation leptons 
negligible. That is, $\beta^\ell = 0$ and a non-zero value for $\alpha^\ell$.

Using these simplifying assumptions in Eq.~\eqref{eq:U1matching}, we obtain the 
following couplings for the $U_1$ leptoquark with the left-handed fermions:
\begin{equation}
  x_L = \frac{M_U}{\Lambda}\sqrt{\frac{-C}{1+|\alpha^\ell|^2}} \begin{pmatrix}
    0\quad{} & 0\quad{} & 0 \\ \alpha^\ell\quad{} & 0\quad{} & 1 \\ \alpha^\ell\quad{} & 0\quad{} & 1
  \end{pmatrix} = \begin{pmatrix}
    0\quad{} & 0\quad{} & 0 \\ x_1\quad{} & 0\quad{} & x_3 \\ x_1\quad{} & 0\quad{} & x_3    
  \end{pmatrix}\,.
\end{equation} 

In this model, the interactions of the $U_1$ leptoquark with fermions are 
governed by just two couplings, $x_1$ and $x_3$. Their numerical values, 
assuming again a leptoquark mass of $M_U = 1.5~\mathrm{TeV}$ and the best fit 
values for $C$ and $\alpha^\ell$ in Scenario II given in Table~\ref{tab:rotation}, are
\begin{equation}
  x_1 = 0.0378\,,\qquad\qquad x_3 = 0.540\,.
\end{equation}

\section{Conclusions}
\label{conclusions}

In this paper, we present the results of the global fit to the flavour
physics observables that exhibit some discrepancies with respect to the
SM values, by considering the NP effects on the Wilson coefficients of
the SMEFT Lagrangian. The global fit includes the $b\to s \mu\mu$
observables; i.e. the Lepton Flavour Universality
ratios $\RKp$, the angular observables $P_5'$, the branching ratio of
$B_s \to \mu\mu$ and all the available data on angular observables in
$B\to K^{(*)} \mu^+ \mu^-$ decay, as well as the $\RDp$ observable, the 
relevant data related to $B\to K^{(*)} e^+ e^-$ decays and
the angular observables measured in different bins for
$B_{s} \to \phi \mu\mu$ decays, $b \to s \nu \bar{\nu}$ and electroweak precision
observables ($W$ and $Z$ decay widths and branching ratios to leptons).
We choose two scenarios in which the condition $C_1=C_3$ is imposed in
order to avoid unwanted contributions to the $B \to K^{(*)} \nu \bar{\nu}$
decays. In Scenario I we fix parameters by assuming that the mixing in 
the first generation is negligible, as already considered
in~\cite{Feruglio:2017rjo}. Scenario II includes non-negligible mixings to the first
generation, allowing us to check the validity of the above assumption.
We found that the better fit is obtained for Scenario II, with
a pull of 6.70 $\sigma$ with respect to the Standard Model, 3.77 $\sigma$
with respect to Scenario I (Table~\ref{tab:rotation}). 
Simultaneous explanation of the \RKp and \RDp anomalies have been also
found in Scenario II (Figure~\ref{im:obs_scbII} and
Table~\ref{tab:observables_rot}).

We show that the Gaussian approximation to characterize the fit is not
successful (see Figure~\ref{im:alphabeta}) and therefore, we use for
the first time in the context of the so-called $B$-anomalies a Machine-Learning Montecarlo
analysis to extract the confidence intervals and correlations between observables.
We found that our procedure reproduce the results obtained in
Section~\ref{sec:Fits} for both the $\Delta
\chi^2$ distribution and the analysis of the impact of each parameter on the global fit.
We also have checked the agreement between the results obtained by the Machine Learning
Montecarlo algorithm proposed in this work and the ones obtained by following the RG equations.
Therefore, we conclude that machine learning, jointly
with the SHAP (SHAPley Additive exPlanation) values, constitute a
suitable strategy to use in this kind of analysis.

This is a promising area of study even if present uncertainties
do not allow us to conclusively establish the presence of physics beyond
the SM, and further analyses are needed.
An observation of the $B^+\to K^+\nu\bar{\nu}$ decay in the near future at Belle II could provide further insight in the $\RKp$ and $\RDp$ anomalies, especially if the excess in the current world average is confirmed. This, together with the expected improved measurements of the electroweak observables in the future linear colliders that we previously studied in~\cite{Alda:2020okk,Alda:2021ruz}, underlines the fundamental role of global analyses and experimental precision in the quest for an explanation of the $B$ anomalies. 

\section*{Acknowledgments}

The authors want to thank Paride Paradisi for useful discussions. The work of J.~A. and S.~P. is partially supported by Spanish grants 
MINECO/FEDER grant FPA2015-65745-P, PGC2018-095328-B-I00 
(FEDER/Agencia estatal de investigaci{\'o}n) and DGIID-DGA No. 2015-E24/2.
J.~A. is also supported by the 
Departamento de Innovaci\'on, Investigaci\'on y Universidad of Arag\'on
goverment, Grant No. DIIU-DGA and the Programa Ibercaja-CAI  de Estancias de Investigaci\'on, Grant No. CB 5/21. 
J.G. has been suported by MICIN under projects PID2019-105614GB-C22 and 
CEX2019-000918-M of ICCUB (\textit{Unit of Excellence Mar{\'\i}a de Maeztu 2020-2023})
and AGAUR (2017SGR754). J.~A. thanks the warm hospitality of the Universit\`a degli Studi di Padova and Istituto Nazionale di Fisica Nucleare during the completion of this work.\\

\appendix
\section{Pulls of the observables in Scenario II}
In this appendix we collect the list of all observables that contribute to the global fit, as
well as their prediction in Scenario II and their pull in both scenario
II (NP pull) and SM (SM pull). Observables are ordered according to their SM
pull, and color-coded according to the difference between the scenario
II and SM pulls: green observables have a better pull in scenario II, 
red observables have a better pull in the SM and white observables have 
a similar pull in both cases.

Predictions for dimensionful observables 
are expressed in the corresponding power of GeV (for example, $\Delta
M_s$ in GeV and $\sigma^0_\mathrm{had}$ in $\mathrm{GeV}^{-2}$). 
The notation $\langle\cdot\cdot\cdot\rangle$ means that the observable
is binned in the invariant mass-squared of the di-lepton system $q^2$, 
with the endpoints of the bin in $\mathrm{GeV}^2$ given in the superscript.
Accordingly, the notation $\frac{\langle \mathrm{BR}\rangle}{\mathrm{BR}}$
denotes a binned branching ratio normalised to the total branching ratio.

\label{app:pulls_scbII}
{\scriptsize 
\begin{longtable}{|c|c|c|c|c|}\hline
 & Observable &	 NP prediction &	 NP pull & SM pull\endhead\hline
0 &	 $\langle \frac{d\overline{\mathrm{BR}}}{dq^2} \rangle(B_s\to \phi \mu^+\mu^-)^{[2.5,\  4.0]}$ &	 $4.5857\times 10^{-8}$ &	 \cellcolor{green!38}3.1 $ \sigma$ &	 3.9 $ \sigma$ \\ \hline
1 &	 $a_\mu$ &	 0.0011659 &	 4.6 $ \sigma$ &	 4.6 $ \sigma$ \\ \hline
2 &	 $R_{\tau \ell}(B\to D^{\ast}\ell^+\nu)$ &	 0.2898 &	 \cellcolor{green!50}0.41 $ \sigma$ &	 3.3 $ \sigma$ \\ \hline
3 &	 $\langle P_2\rangle(B^0\to K^{\ast 0}\mu^+\mu^-)^{[0.1,\  0.98]}$ &	 -0.131 &	 \cellcolor{green!1}3.3 $ \sigma$ &	 3.3 $ \sigma$ \\ \hline
4 &	 $\langle \frac{d\overline{\mathrm{BR}}}{dq^2} \rangle(B_s\to \phi \mu^+\mu^-)^{[1.1,\  2.5]}$ &	 $4.9665\times 10^{-8}$ &	 \cellcolor{green!31}2.4 $ \sigma$ &	 3 $ \sigma$ \\ \hline
5 &	 $\langle \frac{d\overline{\mathrm{BR}}}{dq^2} \rangle(B_s\to \phi \mu^+\mu^-)^{[4.0,\  6.0]}$ &	 $4.8464\times 10^{-8}$ &	 \cellcolor{green!43}2.3 $ \sigma$ &	 3.2 $ \sigma$ \\ \hline
6 &	 $\langle F_L\rangle(B^+\to K^{\ast +}\mu^+\mu^-)^{[2.5,\  4]}$ &	 0.77058 &	 \cellcolor{green!7}3.2 $ \sigma$ &	 3.3 $ \sigma$ \\ \hline
7 &	 $\langle R_{\mu e} \rangle(B^\pm\to K^\pm \ell^+\ell^-)^{[1.1,\  6.0]}$ &	 0.83438 &	 \cellcolor{green!50}0.28 $ \sigma$ &	 3.2 $ \sigma$ \\
\hline
\end{longtable}}

\end{document}